\documentclass[useAMS,usenatbib]{article}
\def\bSig\mathbf{\Sigma}

\newcommand{\ci}{\perp\!\!\!\perp}
\newcommand\BibTeX{{\rmfamily B\kern-.05em \textsc{i\kern-.025em b}\kern-.08em
T\kern-.1667em\lower.7ex\hbox{E}\kern-.125emX}}

\usepackage[latin1]{inputenc}

\usepackage{amsmath}
\usepackage{amssymb}
\usepackage{latexsym}
\usepackage{epsfig}
\usepackage{amsfonts}
\usepackage{moreverb}
\usepackage{natbib}
\usepackage{multirow}
\usepackage[pdftex,colorlinks,bookmarksopen,bookmarksnumbered,citecolor=red,urlcolor=red]{hyperref}
\usepackage{tikz}
\usetikzlibrary{trees}
\usepackage[figuresright]{rotating}
\usepackage{geometry}
 \usepackage[labelfont=bf, labelsep=newline, justification=centering, textfont=it]{caption} 
\title{Data-Driven Confounder Selection via Markov and Bayesian Networks}

\author{Jenny H\"aggstr\"om$^{\ast}$\\
	   Department of Statistics, USBE, Ume{\aa} University, SE-901 87 Ume{\aa}, Sweden }

\begin{document}


\date{}





%


\maketitle
\vspace{-5ex}
\begin{center}${}^{\ast}$\small{\textit{email}: jenny.haggstrom@umu.se}\end{center}

%

\begin{abstract}
\noindent
\textsc{Summary:} To unbiasedly estimate a causal effect on an outcome unconfoundedness is often assumed. If there is sufficient knowledge on the underlying causal structure then existing confounder selection criteria can be used to select subsets of the observed pretreatment covariates, $X$, sufficient for unconfoundedness, if such subsets exist. Here, estimation of these target subsets is considered when the underlying causal structure is unknown. The proposed method is to model the causal structure by a probabilistic graphical model, e.g., a Markov or Bayesian network, estimate this graph from observed data and select the target subsets given the estimated graph. The approach is evaluated by simulation both in a high-dimensional setting where unconfoundedness holds given $X$ and in a setting where unconfoundedness only holds given subsets of $X$. Several common target subsets are investigated and the selected subsets are compared with respect to accuracy in estimating the average causal effect. The proposed method is implemented with existing software that can easily handle high-dimensional data, in terms of large samples and large number of covariates. The results from the simulation study show that, if unconfoundedness holds given $X$, this approach is very successful in selecting the target subsets, outperforming alternative approaches based on random forests and LASSO, and that the subset estimating the target subset containing all causes of outcome yields smallest MSE in the average causal effect estimation.
\end{abstract}

\section{Introduction}
\label{Sec:1}
To get an unbiased estimate of a causal effect, of a treatment on some outcome, the treatment assignment is often assumed to be unconfounded, which is the case, e.g.,  when assignment to treatment is randomized. In an observational study treatment assignment is not randomized, and to get unbiased causal effect estimates we need to make sure that the assumption of unconfoundedness is plausible when conditioning on some set of covariates. Two important questions are: 1) Which set of covariates should we aim to condition on? and 2) How should we in practice go about to select the latter set of covariates? Often the answer to the first question has been "the set of covariates that are common causes of treatment and outcome" or "all observed pretreatment covariates", hereinafter referred to as 'the common cause criterion' and 'the pretreatment criterion', respectively. However, in response to \cite{DR:07}, in a series of letters to the editor and author's replies (\citeauthor{IS:08}, \citeyear{IS:08}; \citeauthor{DR:08}, \citeyear{DR:08}; \citeauthor{JP:09a}, \citeyear{JP:09a}; \citeauthor{AS:09}, \citeyear{AS:09}; \citeauthor{DR:09}, \citeyear{DR:09}) and later on in \cite{VS:11} it was discussed under what circumstances conditioning on the covariate set defined by the common cause criterion or the pretreatment criterion will in fact induce bias in the causal effect estimate instead of reducing it. 

In an attempt to mediate between the standpoints of Pearl, Shrier, Sj\"olander and Rubin \cite{VS:11} proposed an alternative covariate selection criterion, 'the disjunctive cause criterion'. The disjunctive cause criterion entails selecting all covariates that are causes of treatment and/or causes of outcome, and will, under certain assumptions, result in unconfoundedness. Specifically, the disjunctive cause criterion is similar to 'the backdoor path criterion' \citep{JP:95} in the sense that the covariate set selected by the disjunctive cause criterion will suffice to block all backdoor paths from treatment to outcome if such a set exists. The main practical difference between these two criteria is that to use the backdoor path criterion, knowledge of the full causal structure of the data is needed but to use the disjunctive cause criterion it is sufficient to know which covariates are causes of the treatment and which covariates are causes of the outcome. \cite{VS:11} dismiss the practical usefulness of the backdoor path criterion and point out that "In a number of analyses in the biomedical and social sciences, such complete knowledge of causal structures is unlikely" (Section 1, p.1406). Regarding the knowledge of the causal structure that is required for use of the disjunctive cause criterion, they are however optimistic, stating that "In many epidemiological and biomedical applications, subject matter experts have intuitive knowledge of whether each covariate is a cause of the treatment or the outcome" (Section 6, p.1411). If it is indeed the case, that we have the knowledge required to use the disjunctive cause criterion (or the backdoor path criterion), then the problem of covariate selection, with respect to unconfoundedness, is solved and the question in 2) is redundant. 

However, when knowledge of the causal structure is not sufficient for use of the disjunctive cause criterion the process of covariate selection can be aided by data-driven procedures. Moreover, even in cases where the disjunctive cause criterion can be used, mean squared error of nonparametric estimators of causal effects may be improved by further reducing the dimensionality of the covariate set \citep*{dLWR:11}. Given a set of covariates such that conditioning on this set unconfoundedness is upheld the latter authors propose general algorithms, in line with the common cause criterion, for selecting minimal sets of covariates such that unconfoundedness still holds when conditioning on the selected sets. These algorithms are implemented using marginal coordinate hypothesis testing (continuous covariates) and kernel smoothing (continuous and/or discrete covariates) in the \texttt{R} package \texttt{CovSel} \citep*{CovSel}. These implementations have in simulation studies \citep*{PHWdL:13} been shown to perform well in reducing the dimensionality of the covariate set while still upholding unconfoundedness, thereby resulting in improved mean squared error. For other recently proposed covariate selection procedures see, e.g., \cite{PHWdL:13}, \cite{SLG:15} and references therein. Existing proposals have in common that they are computer intensive, yielding prohibitive running times in high dimensional applications (in terms of number of covariates and number of units).

In this paper we propose and study the use of Markov and Bayesian network algorithms, more precisely Max-Min Parents and Children (MMPC) and Max-Min Hill-Climbing (MMHC) \citep*{TI:06}, in conjunction with the covariate selection algorithms in \cite{dLWR:11}. MMHC has in empirical evaluations been shown to outperform the PC algorithm \citep{SGS:00}, previously studied  by \cite*{MKB:09} in a similar causal inference setting (although not for explicitly selecting covariates), both with regard to computation time and in ability to accurately estimate the true causal structure \citep{TI:06}. To the author's knowledge, using estimated graphs to explicitly select covariates to control for, as a step completely separated from the nonparametric estimation of causal effects, has not been proposed and studied elsewhere. 

The performance of the proposed data-driven covariate selection procedures is investigated, using simulations, in two general high-dimensional scenarios: 1) Unconfoundedness holds given the full covariate set, 2) Unconfoundedness does not hold given the full covariate set, but it does hold given a subset of the full covariate set. In scenario 1), the results show that this approach is very successful in selecting the target covariate subsets. Furthermore, targeting the subset containing all causes of outcome often yields smallest MSE in the average causal effect (ACE) estimation, but the magnitude of the reduction in MSE (relative to conditioning on the full covariate set) depends on the ACE estimator. The proposed covariate selection algorithms are implemented in the \texttt{R} package \texttt{CovSelHigh} \citep{CovSelHigh}.

The remainder of this paper is organized as follows. In Section \ref{Sec:2} relevant notation and concepts from causal inference are reviewed. Section \ref{Sec:3} focuses on covariate selection when the causal structure is known and Section \ref{Sec:4} on covariate selection using Markov and Bayesian network algorithms when the causal structure is unknown. 
In Section \ref{Sec:5} the simulation study is  presented. In Section \ref{Sec:6} the proposed approach is illustrated using a large register data set with which the ACE of C-section delivery on asthma medication early in life is estimated. The paper is concluded with a discussion in Section \ref{Sec:7}.

\section{Context and Terminology}\label{Sec:2}
\subsection{Potential Outcomes and Unconfoundedness}\label{Sec:2.1}

Let $T$ denote a binary treatment, $Y$ denote outcome and $X$ denote the set of observed pretreatment covariates, i.e., the full covariate set. 
Within the potential outcome framework  (\citeauthor{JN:23}, \citeyear{JN:23}; \citeauthor{DN:74}, \citeyear{DN:74}) we let $Y(1)$ and $Y(0)$ denote the potential outcomes for $Y$ under the two treatments $T=1$ and $T=0$, respectively. Since only one treatment assignment is possible for each unit only one of the two potential outcomes is observed, $Y=Y(0)(1-T)+Y(1)T$. In this paper, the ACE, $\beta=E\{Y(1)-Y(0)\}$, is the parameter of interest.

If treatment assignment is not randomized, $\beta$ is identified if a unit's potential outcomes does not depend on the treatments received by other units (stable unit treatment assumption, SUTVA) \citep{DR:90} and we have available a set of pretreatment covariates $S$ such that the probability of receiving either treatment conditional on $S$ is bounded away from 0 (overlap assumption) and such that the treatment assignment is unconfounded conditional on $S$. Letting $\ci$ mean "is independent of" \citep{PD:79} the assumption of unconfoundedness is upheld if $Y(1), Y(0)  \ci T \mid S.$

\subsection{Graphical Models}\label{Sec:2.2}
A graph $G = (V,E)$ consists of a set of vertices $V = \{V_i: i\in 1, \ldots, p\}$ and a set of edges $E$. The vertices represent random variables, and the edges describe pairwise relationships among the variables. Here, the two types of graphs considered are undirected graphs and directed acyclic graphs (DAGs). In an undirected graph all edges are undirected (---) and in a directed graph all edges are directed $(\rightarrow)$. A directed graph with no cycles is a DAG. Two vertices are adjacent if they are connected by an edge and the neighbors of a vertex $V_i$ consists of all vertices adjacent to $V_i$. Furthermore, we define a path in a graph as a sequence of edges connecting vertices such that each vertex on the path is visited only once. In a DAG a vertex is a collider on a path if the path enters and leaves the vertex via arrowheads. In this paper we interpret directed edges as causal, i.e., $V_1 \rightarrow V_2$ means that $V_1$ is a cause of $V_2$, also $V_1$ is said to be a parent of $V_2$ and $V_2$ the child $V_1$ \citep{JP:09}. In an undirected graph the Markov blanket of a vertex consists of the vertex's neighbors and in a DAG the Markov blanket of a vertex consists of the vertex's parents, children and children's other parents. The local Markov property holds with respect to an undirected graph $G$ if there exists a joint probability distribution for $V$ such that, for each $V_i \in V$, $V_i$ is conditionally independent of its non-neighbors given its neighbors. For a DAG the local Markov property holds if there exists a joint probability distribution for $V$ such that, for each $V_i \in V$, $V_i$ is conditionally independent of its non-descendants given its parents.

A Markov network is a model consisting of an undirected graph, $G = (V,E)$, and a joint probability distribution $P$ defined over $V$ such that the local Markov property holds with respect to $G$. Similarly, a Bayesian network is a model consisting of a  DAG, $G = (V,E)$, and a joint probability distribution $P$ defined over $V$ such that the local Markov property holds with respect to $G$. A probability distribution is faithful to the graph if it obeys no further conditional independence relations than what are entailed by the local Markov property. 

 \section{Target Covariate Subsets}\label{Sec:3}
 In this section, target covariate subsets are defined using the languages of potential outcomes and graphical models. Let $G_t=(V_t, E_t)$ with $V_t=\{X, T, Y(t)\}$ for $t=0,1$. The subset of $X$ that includes all causes of treatment is defined as $X_{\rightarrow T}=\{X_i \in X : X_i \rightarrow T \in E_0 \cup E_1\}$. Similarly, let $X_{\rightarrow Y}=X_{\rightarrow Y}^0 \cup X_{\rightarrow Y}^1=\{X_i \in X : X_i \rightarrow Y(0) \in E_0\}\cup \{X_i \in X : X_i \rightarrow Y(1) \in E_1\}$ be the subset of $X$ that includes all causes of outcome. Furthermore, let the subset of $X_{\rightarrow T}$ consisting of elements dependent with outcome be defined as $Q_{\rightarrow T}=Q_{\rightarrow T}^0 \cup Q_{\rightarrow T}^1$ where $Q_{\rightarrow T}^t \subseteq  X_{\rightarrow T}$ such that $Y(t)  \ci  X_{\rightarrow T} \setminus  Q_{\rightarrow T}^t \mid  Q_{\rightarrow T}^t$,  for $t=0, 1$, and let the subset of $X_{\rightarrow Y}$ consisting of elements dependent with treatment be defined as $Z_{\rightarrow Y}=Z_{\rightarrow Y}^0 \cup Z_{\rightarrow Y}^1$ where $ Z_{\rightarrow Y}^t \subseteq  X_{\rightarrow Y}^t$ such that $T \ci  X_{\rightarrow Y}^t \setminus  Z_{\rightarrow Y}^t \mid  Z_{\rightarrow Y}^t$, for $t=0, 1$.  Defined in line with the common cause criterion, $Q_{\rightarrow T}$ is a subset of the covariates that cause treatment, and $Z_{\rightarrow Y}$  a subset of the covariates that cause outcome.
 
The following, similar but not identical to the above, sets are defined in \cite{dLWR:11}: $X_T\subseteq  X$ such that $T \ci  X \setminus  X_T \mid  X_T$, $X_Y=X_0 \cup X_1$ where $X_t \subseteq X$ such that $Y(t) \ci  X \setminus  X_t \mid  X_t$ for $t=0, 1$,  $Q_t \subseteq X_T$ such that $Y(t) \ci  X_T \setminus  Q_t \mid  Q_t$ for $t=0, 1$ and $Z_t \subseteq X_Y$ such that $T \ci  X_Y \setminus  Z_t \mid  Z_t$ for $t=0, 1$. As long as we have unconfoundedness given $X$ the following equalities hold: $X_{\rightarrow T}=X_T$, $X_{\rightarrow Y}=X_Y$, $Q_{\rightarrow T}=Q$ and $Z_{\rightarrow Y}=Z$. The latter authors also state assumptions under which these sets are unique \citep[][Lemmas A2-A5]{dLWR:11} and $Z$ and $Q$ are minimal sets that cannot be reduced without violating unconfoundedness \citep[][Proposition 8]{dLWR:11}. The above equalities do not hold in general since $X_{\rightarrow T}$ and $ X_{\rightarrow Y}$ are defined in terms of edges present in a DAG while \cite{dLWR:11} define all sets in terms of conditional independencies.

The subset of $X$ that includes all causes of treatment and/or outcome, i.e., the disjunctive cause criterion subset, is defined as $X_{\rightarrow T, Y}=X_{\rightarrow T}\cup X_{\rightarrow Y}$.  \cite{VS:11} suggest a criterion where covariates unassociated with the outcome are iteratively discarded from $X_{\rightarrow T, Y}$, henceforth 'the disjunctive cause criterion with backward selection', and the subset of $X_{\rightarrow T, Y}$ consisting of elements that are associated with outcome is defined as $W_{\rightarrow Y}=W_{\rightarrow Y}^0 \cup W_{\rightarrow Y}^1$ where $W_{\rightarrow Y}^t \subseteq  X_{\rightarrow T , Y}$ such that $Y(t)  \ci  X_{\rightarrow T, Y} \setminus  W_{\rightarrow Y}^t \mid  W_{\rightarrow Y}^t$,  for $t=0, 1$. If $X_{\rightarrow T}$ and $X_{\rightarrow Y}$ are uniquely defined then it follows that $X_{\rightarrow T, Y}$ is unique and under assumptions similar to Lemma  A3 in \cite{dLWR:11} so is  $W_{\rightarrow Y}$.

For illustrative purposes consider the causal diagram in Figure \ref{Fig:1}. Here, the observed pretreatment covariates $X=\{X_i:i\in 1, \ldots, 10\}$ are the only variables affecting $T$ and $Y$ and thus all of the covariate selection criteria mentioned in Section \ref{Sec:1} would result in unconfoundedness. The target sets are $X_{\rightarrow T}=X_T=\{X_1, X_2, X_3, X_4, X_7\}$, $X_{\rightarrow Y}=X_Y=\{X_1, X_2, X_5, X_6, X_8\}$, $Q_{\rightarrow T}=Q=\{X_1, X_2, X_7\}$, $Z_{\rightarrow Y}=Z=\{X_1, X_2, X_8\}$, $X_{\rightarrow T, Y} =\{X_1, X_2, X_3, X_4, X_5, X_6, X_7, X_8\}$ and $W_{\rightarrow Y}=\{X_1, X_2, X_5, X_6, X_7, X_8\}$.

The common cause criterion would select $S=\{X_1, X_2, X_7\}=Q_{\rightarrow T}$, the pretreatment criterion would select $S=X$, the backdoor path criterion would select $S=\{X_1, X_2, X_3, X_4, X_7\}$, the disjunctive cause criterion would select $S=X_{\rightarrow T, Y} $ and the disjunctive cause criterion with backward selection would select $S=W_{\rightarrow Y}$.

 \begin{figure}
\begin{center}
\resizebox{0.5\textwidth}{!}{
  \begin{tikzpicture}
   
           \begin{scope}[font=\small]
               \draw [->] (-0.25,-0.4) -- (-0.25,-0.8);
         \draw(-0.25,-1.0) node {$X_1$};
           \draw [->] (2.55,-0.9) -- (3.25,-1.6);
           
             \draw(-0.25,-0.2) node {$X_2$};
  \draw [->] (-0.7,-0.5) -- (-1.5,-1.5);
    \draw [->] (0.1,-0.5) -- (0.8,-1.6);
    \draw [->] (-0.7,-1.1) -- (-1.3,-1.6);
    \draw [->] (0.1,-1.1) -- (0.6,-1.7);
  \draw(-1.75,-2) node {$T$};
     \draw(-4.4,-2) node {$X_3$};
   \draw [->] (-4.0,-2) -- (-2.1,-2); 
       \draw(1,-2) node {$Y(t)$};
      \draw(3.65,-2) node {$X_5$};
  \draw [<-] (1.35,-2) -- (3.25,-2);

          \draw(2.25,-0.5) node {$X_6$};  
     \draw [->] (2.1,-0.8) -- (1.4,-1.6);   
       \draw(0.4,0.3) node {$X_8$};  
     \draw [->] (0.5,-0.1) -- (0.9,-1.5);
        \draw(-1.2,0.3) node {$X_7$};  
   \draw [->] (-1.2,-0.1) -- (-1.7,-1.5);
    \draw [->] (-0.8,0.3) -- (0.0,0.3);
    \draw(-0.5,-3.0) node {$X_9$};
                    \draw(-3.0,0.0) node {$X_{10}$};   
                      \draw(-3,-4.0) node {$X_4$};
  \draw [->] (-2.9,-3.7) -- (-2.0,-2.3);

  \end{scope}

  \end{tikzpicture}
}
\end{center}
\caption{A causal DAG without unobservables.  }
\label{Fig:1}
\end{figure}

Now consider the causal diagram in Figure \ref{Fig:2}. Here, in addition to $T$, $Y$ and $X=\{X_i:i\in 1, \ldots, 10\}$ we have a set of unobserved variables $U=\{U_1, U_2, U_3\}$ affecting $T$ and/or $Y$. The target sets $X_{\rightarrow T}$, $X_{\rightarrow Y}$, $Z_{\rightarrow Y}$, and $X_{\rightarrow T, Y}$ remain as for Figure \ref{Fig:1} but now $Q_{\rightarrow T}=\{X_1, X_2, X_4, X_7 \}$, and $W_{\rightarrow Y}=\{X_1, X_2, X_4, X_5, X_6, X_7, X_8\}$.

As pointed out by \cite{VS:11}, in this setting not all of the above covariate selection criteria would result in unconfoundedness. Except for the disjunctive cause criterion with backward selection, all of the criteria would in this setting select the same sets as in the previous setting. However, the set selected by the common cause criterion,  $\{X_1, X_2, X_7\}\neq Q_{\rightarrow T}$, would not result in unconfoundedness since it does not include the covariate $X_4$, which is now related to both $T$ and $Y$, the latter through the unobserved variable $U_3$. 
The set selected by the pretreatment criterion would fail to achieve unconfoundedness due to the inclusion of the covariate $X_9$, which is now a collider on the path between $T$ and $Y$ due to the unobserved variables $U_1$ and $U_2$. Conditioning on $X_9$ will thus open up this path between $T$ and $Y$ and introduce the so called $M$-bias \citep{SG:03}. The sets selected by the backdoor path criterion and the disjunctive cause criterion would however achieve unconfoundedness since both sets would include $X_4$ but not $X_9$. The disjunctive cause criterion with backward selection would select $S=\{X_1, X_2, X_4, X_5, X_6, X_7, X_8\}=W_{\rightarrow Y}$ which includes $X_4$ since it is now associated with the outcome and conditioning on this set upholds unconfoundedness. Conditioning on $Q_{\rightarrow T}$ would also, in this case, result in unconfoundedness.

Note that here the sets defined in \cite{dLWR:11} are $X_T=X_{\rightarrow T}\cup \{X_9\}$, $X_Y=X_{\rightarrow Y}\cup \{X_4, X_9\}$, $Q=Q_{\rightarrow T}\cup \{X_9\}$ and $Z=Z_{\rightarrow Y}\cup \{X_4, X_9\}$, all including $X_9$. 
 
\begin{figure}

\begin{center}
\resizebox{0.5\textwidth}{!}{
  \begin{tikzpicture}
   
           \begin{scope}[font=\small]
                          \draw [->] (-0.25,-0.4) -- (-0.25,-0.8);
        \draw(-0.25,-1.0) node {$X_1$};
           \draw [->] (2.55,-0.9) -- (3.25,-1.6);
             \draw(-0.25,-0.2) node {$X_2$};
  \draw [->] (-0.7,-0.5) -- (-1.5,-1.5);
    \draw [->] (0.1,-0.5) -- (0.8,-1.6);

    \draw [->] (-0.7,-1.1) -- (-1.3,-1.6);
    \draw [->] (0.1,-1.1) -- (0.6,-1.7);
  \draw(-1.75,-2) node {$T$};
     \draw(-4.4,-2) node {$X_3$};
   \draw [->] (-4.0,-2) -- (-2.1,-2); 
       \draw(1,-2) node {$Y(t)$};
       \draw(3.65,-2) node {$X_5$};
  \draw [<-] (1.35,-2) -- (3.25,-2);     
      \draw(2.25,-0.5) node {$X_6$};  
     \draw [->] (2.1,-0.8) -- (1.4,-1.6);   
       \draw(0.4,0.3) node {$X_8$};  
     \draw [->] (0.5,-0.1) -- (0.9,-1.5);
        \draw(-1.2,0.3) node {$X_7$};  
   \draw [->] (-1.2,-0.1) -- (-1.7,-1.5);
    \draw [->] (-0.8,0.3) -- (0.0,0.3);
    \draw(-0.5,-3.0) node {$X_9$};
      \draw(-1.75,-3.0) node {$U_1$};
        \draw [->] (-1.75,-2.7) -- (-1.75,-2.3);
        \draw [->] (-1.55,-3.0) -- (-0.8,-3.0);
            \draw(-0.5,-4.0) node {$U_2$};   
                       \draw [->] (-0.5,-3.7) -- (-0.5,-3.3);
                         \draw [->] (-0.3,-3.7) -- (0.8,-2.3);                  
                    \draw(-3.0,0.0) node {$X_{10}$};   
                      \draw(-3,-4.0) node {$X_4$};
  \draw [->] (-2.9,-3.7) -- (-2.0,-2.3);
    \draw(1,-5.0) node {$U_3$};
        \draw [<-] (-2.7,-4.0) -- (0.7,-4.9);  
        \draw [->] (1,-4.7) -- (1,-2.4);

  \end{scope}

  \end{tikzpicture}
}
\end{center}
\caption{A causal DAG with unobservables. }
\label{Fig:2}
\end{figure}

\section{Covariate Selection When the Causal Structure is Unknown}\label{Sec:4}
Given the setup stated in Section \ref{Sec:2}, and no further knowledge on the causal structure, only the pretreatment criterion can be readily used without aid of data-driven procedures. If we, in some way, from data estimate the dependence structure in the form of an undirected or directed graph then we can use the estimated graph to select covariates by reading off which covariates are related to $T$ and/or $Y \mid T=t$ for $t=0,1$. 

There are many different methods available for estimating Markov and Bayesian networks \citep[see, e.g., ][and references therein]{FNP:99, SGS:00,DC:02, TI:06}. In this paper, the Max-Min Parents and Children Algorithm (MMPC) and the Max-Min Hill-Climbing Algorithm (MMHC) are used to estimate the underlying structure of the data.  Algorithms used for estimating such networks can be classified as either constraint-based or score-based. MMHC is a hybrid algorithm which as a first step uses the constraint-based MMPC algorithm to estimate a Markov network, i.e., an undirected graph, and as a second step uses the score-based local optimization technique hill-climbing (similar to steepest ascent) to find the Bayesian network, i.e., the DAG, that best fits the data. For the purpose of estimating the graphs we assume that we have a copy of the data $\{X, T, Y\}$ where any continuous variables have been discretized. 

\subsection{The Max-Min Parents and Children Algorithm}
MMPC estimates the underlying graph structure by testing if the conditional independencies between the variables implied by a Markov network hold. One commonly used conditional independence test is based on the information-theoretic measure mutual information. Using the observed frequencies for variables $V_i$, $V_j$ and all the configurations of the variables in the conditioning set MI is estimated by $\widehat{\mbox{MI}}(V_i, V_j \mid V_k=\{V \setminus \{V_i, V_j\}\})=n^{-1}\sum_{abc}n_{ijk}^{abc}\mbox{ln}\{n_{ijk}^{abc}n_{k}^{c}(n_{ik}^{ac}n_{jk}^{bc})^{-1}\}$, where $n_{ijk}^{abc}$ is the size of the subsample where $V_i=a$, $V_j=b$ and $V_k=c$. $n_{k}^{c}$, $n_{ik}^{ac}$ and $n_{jk}^{bc}$ are defined analogously. $\widehat{\mbox{MI}}$ is proportional to the likelihood-ratio test (by a factor of $2n$) and is asymptotically $\chi^2$-distributed with $(A-1)(B-1)C$ degrees of freedom, where $A$, $B$ and $C$ are the number of distinct configurations of $V_i$, $V_j$ and $V_k$.

The goal of MMPC is, for each variable $V_i$, $i=1,\ldots, p$, to return the set containing the variable's neighbors, i.e., the variable's Markov blanket, $MB^i$. For the variable $V_i$ and supposing that the rest of the variables in the graph are ordered as $(V_1, \ldots, V_J)$ MMPC starts in phase 1 with the empty set as the candidate $MB^i$, $CMB_0^i=\emptyset$, and updates the candidate $MB^i$ as follows, $j=1,\ldots,J$,

\begin{equation*}
CMB_j^i = \left\{
\begin{array}{ll}
CMB_{j-1}^i & \text{if } V_j \ci V_i\mid CMB_{j-1}^i,\\
CMB_{j-1}^i\cup\{V_j\}& \text{otherwise}.
\end{array} \right.
\end{equation*}

In phase 2 MMPC starts with the set $CMB_J^i$ from phase 1 as the candidate $MB^i$, i.e., $CMB_0^i=CMB_J^i$. Suppose that the variables in the set $CMB_J^i$ are ordered as $(V_1, \ldots, V_K)$ then the set is updated as follows, $k=1, \ldots, K$,

\begin{equation*}
CMB_k^i = \left\{
\begin{array}{ll}
{CMB}_{k-1}^i\setminus \{V_k\} & \text{if } \exists A \subseteq CMB_{k-1}^i\setminus\{V_k\} \\
&\text{s.t. }V_i\ci V_k \mid A,\\
CMB_{k-1}^i& \text{otherwise}.
\end{array} \right.
\end{equation*}

After running phase 1 and phase 2 for all $V_i$, $i=1, \ldots, p$, an attempt to remove any variables included in the final CMBs that are not a neighbor (false positives) is made in phase 3. Start with the set $CMB_0^i=CMB_K^i$ and suppose that the variables in the set $CMB_K^i$ are ordered as $(V_1, \ldots, V_L)$ then the set is updated as follows, $l=1, \ldots, L$,

\begin{equation*}
CMB_l^i = \left\{
\begin{array}{ll}
CMB_{l-1}^i & \text{if } V_i \in CMB_{K}^l \\
CMB_{l-1}^i\setminus \{V_l\}& \text{otherwise}.
\end{array} \right.
\end{equation*}

The final set for variable $V_i$ is $MB^i=CMB_L^i$. With the knowledge of all neighbors of all the variables in the graph the undirected graph can be constructed.

\subsection{The Max-Min Hill-Climbing Algorithm}
In the first step MMPC is performed. In the second step MMHC attempts to identify the Bayesian network that maximizes a score function indicating how well the graph fits the data, e.g., AIC, BIC or similar criteria. An empty graph is the starting point and a new candidate graph is generated by performing one of the following alterations to the current candidate graph: single-edge addition, single-edge deletion, or single-edge direction reversal. The alteration that leads to the largest increase in score is performed. The procedure is iterated until there is no alteration that increases the score. The optimization is constrained to only consider adding an edge if it is present in the undirected graph returned by MMPC in the first step.

\subsection{Estimation of Target Covariate Subsets}
The target covariate sets defined earlier are estimated by fitting a number of discrete Markov or Bayesian networks in a stepwise manner adhering to the covariate selection algorithms in \cite{dLWR:11}. To be clear, we only estimate certain graphs and the sets of covariates in the Markov blankets of these estimated graphs are what is meant by "estimated target covariate sets". Given the assumed temporal order between $X$, $T$ and $Y$ we always incorporate the constraints that $T$ and $Y \mid T=t$ for $t=0,1$ have no children in $X$. Consider estimating the subgraph including only vertices $V=\{X, T\}$ then $\widehat{X}_ {\rightarrow T}$ is defined as the estimated Markov blanket of $T$. Given $\widehat{X}_{\rightarrow T}$ we estimate the subgraphs including only vertices $V=\{\widehat{X}_{\rightarrow T}, Y \mid T=t\}$ for $t=0,1$ and define $\widehat{Q}_{\rightarrow T}=\widehat{Q}_{\rightarrow T}^0\cup\widehat{Q}_{\rightarrow T}^1$ as the union of the estimated Markov blankets for $Y \mid T=t$ for $t=0,1$. Similarly, we estimate $\widehat{X}_{\rightarrow Y}$ by estimating the subgraphs including only vertices $V=\{X, Y \mid T=t\}$ for $t=0,1$ and define $\widehat{X}_{\rightarrow Y}=\widehat{X}_{\rightarrow Y}^0\cup\widehat{X}_{\rightarrow Y}^1$ as the union of the estimated Markov blankets for $Y \mid T=t$ for $t=0,1$. Given $\widehat{X}_{\rightarrow Y}$ we estimate the subgraphs including only vertices $V=\{\widehat{X}_{\rightarrow Y}^t, T\}$ for $t=0,1$ and define $\widehat{Z}_{\rightarrow }=\widehat{Z}_{\rightarrow Y}^0\cup\widehat{Z}_{\rightarrow Y}^1$ as the union of the estimated Markov blankets for $T$. Furthermore, $\widehat{X}_{\rightarrow T, Y}=\widehat{X}_{\rightarrow T}\cup\widehat{X}_{\rightarrow Y}$ and given $\widehat{X}_{\rightarrow T, Y}$ we estimate the subgraphs including only vertices $V=\{\widehat{X}_{\rightarrow T, Y}, Y \mid T=t\}$ for $t=0,1$ and define $\widehat{W}_{\rightarrow Y}=\widehat{W}_{\rightarrow Y}^0\cup\widehat{W}_{\rightarrow Y }^1$ as the estimated Markov blankets for $Y \mid T=t$ for $t=0,1$. Although MMHC results in a DAG, for our purposes, the directionality of the edges give no added information since we assume that $T$ and $Y \mid T=t$, for $t=0,1$, have no children in $X$ and we are not interested in the relations between the covariates. MMHC can however result in a graph with fewer edges than those present in the undirected graph produced by MMPC.

Note that technically this estimation strategy violates Rubin's "no outcome data"-policy (\citeauthor{DR:07}, \citeyear{DR:07}; \citeauthor{DR:08b}, \citeyear{DR:08b}) which entails that study design, e.g., confounder selection, should be performed without any use of outcome data. However, the ACE is never estimated in the confounder selection process and outcome data are only considered separately for each treatment group, thus avoiding any difference in outcome between the treatment groups to influence the confounder selection and subsequent ACE estimation.
\subsection{Theoretical Results}
\begin{itemize}
\item[C1.] $X$, $T$ and $Y(t)$, $t=0,1$ are all discrete random variables.
\item[C2.] $Y(t)$ and $T$ have no children in $X$ and $Y(t)$, $t=0,1$ is not a parent of $T$. All confounders are observed.
\item[C3.] The underlying true causal structures are DAGs, denoted $G_{X, T, Y(t)}$ for $t=0,1$, involving the set of vertices $V_{X, T, Y(t)}=\{X,T,Y(t)\}$.
\item[C4.] There exist joint probability functions, denoted $p_{X, T, Y(0)}$ and $p_{X, T, Y(1)}$, such that the local Markov property holds with respect to $G_{X, T, Y(t)}$, for $t=0,1$.
\item[C5.] $p_{X, T, Y(0)}$ and $p_{X, T, Y(1)}$ are faithful to $G_{X, T, Y(t)}$, for $t=0,1$, respectively.
\item[C6.] A perfect conditional independence oracle is available.
\end{itemize}

\textsc{Theorem 1:} \textit{When conditions} $C1-C6$ \textit{are satisfied, the estimated target covariate sets resulting from using MMPC will equal the true target covariate sets. That is,} $\widehat{X}_ {\rightarrow T}=X_ {\rightarrow T}$, $\widehat{Q}_{\rightarrow T}=Q_{\rightarrow T}$, $\widehat{X}_{\rightarrow Y}=X_{\rightarrow Y}$, $\widehat{Z}_{\rightarrow }=Z_{\rightarrow Y}$, $\widehat{X}_{\rightarrow T, Y}=X_{\rightarrow T, Y}$ \textit{and} $\widehat{W}_{\rightarrow Y}=W_{\rightarrow Y}$.
\vskip 0.25cm
\noindent
A proof of Theorem 1 appear in Appendix A. 
\vskip 0.25cm
\noindent
\textsc{Remark 1:} \textit{Conditions C2-C5 are fairly reasonable and common assumptions in settings like these. If the outcome and/or some of the covariates are continuous variables (C1 violated) and there is a need for discretizing prior to performing confounder selection via MMPC, information will be lost and this can affect the performance of the confounder selection procedure in practice. Most notably, we do not have access to a perfect conditional independence oracle (C6 violated) and hence the quality of the confounder selection procedure will depend on the properties of the method used for determining conditional independencies. }
\vskip 0.25cm
\noindent
As far as the author knows, there is no theoretical results regarding the final output of MMHC. However, since MMHC performs MMPC in the first step, when conditions C1-C6 are satisfied the estimated target covariate sets resulting from using MMHC will be equal to, or subsets of, the estimated target covariate sets resulting from MMPC.

\section{Simulation Study}\label{Sec:5}
A simulation study is performed to evaluate 1) the ability of MMPC and MMHC, respectively, to retrieve the target covariate subsets $X_ {\rightarrow T}$, $Q_{\rightarrow T}$, $X_{\rightarrow Y}$, $Z_{\rightarrow Y}$, $X_{\rightarrow T, Y}$ and $W_{\rightarrow Y}$ and 2) to what extent the retrieved covariate sets result in unconfoundedness and 3) the impact of the selected sets on the estimation of  the ACE. Comparisons are made with two other methods sometimes used for variable selection, namely random forests \citep[RF;][]{LB:01} and LASSO \citep{RT:86}.

\subsection{Simulation design}
All simulations are repeated with 1000 iterations each, with sample sizes $n$ = 500, 1000, 2000, 10000 and 100 covariates included in $X$. Data generation and all computations are performed with the software \texttt{R} \citep{R} using the \texttt{R} package \texttt{CovSelHigh} \citep{CovSelHigh}. 

\subsubsection{Setting 1: Unconfoundedness holds given $X$}\label{Setting1}
 In this setting the core causal structure corresponds to Figure \ref{Fig:1}. In addition to the ten covariates visible in Figure \ref{Fig:1} 90 additional covariates are generated, related to each other but not to the first ten covariates. Hence, the complete covariate set consists of 100 covariates , $ X = \{X_i:i\in 1, ..., 100\}$. A mixture of continuous and discrete covariates are simulated and the ten covariates in the core causal structure are generated according to the following specification: $(R_1,X_2)^T, (X_5, R_2)^T \sim \mbox{N}\big((0,0)^T,((1,0.5)^T(1,0.5)^T)\big),\,\, X_1=I(R_1>0), \,X_6=I(R_2>0)$, $(X_7,X_8)^T\sim \mbox{Bernoulli}\big((0.5,0.5)^T$,\\$((1,0.7)^T(1,0.7)^T)\big),
X_3, \,X_{10} \sim \mbox{Bernoulli}(0.5), \,\,X_4, \,X_9 \sim \mbox{N}(0,1).$

We have that $\text{Corr}(X_1, X_2)=\text{Corr}(X_5, X_6)=0.4$ and $\text{Corr}(X_7, X_8)=0.7$. For a full specification of how the rest of the covariates are generated see the function \texttt{cov.sel.high.sim} in the \texttt{R} package \texttt{CovSelHigh}. Let $f_T(X)=3 -2X_1 -2X_2 -2X_3 -X_4 -2X_7$ and  $f_Y(X)=4X_1+ 2X_2 + 2X_5+ 4X_6 + 4X_8$, then the treatment variable, $T$, is generated from $n$ Bernoulli trials with the treatment probability $\mbox{P}(T=1 \mid  X) = [1 + \exp\{f_T(X)\}]^{-1}$. The coefficients are chosen such that $\mbox{E}(T) = 0.5$. Three outcome models are generated: one linear, one binary and one nonlinear. The linear outcome model, for $t=0,1$, is $Y(t) = 2+2t+f_Y(X) + \varepsilon_t$, where $\varepsilon_t\sim \mbox{N}(0, 1)$. The binary outcome model, for $t=0,1$, is generated as Bernoulli trials with probabilities $\mbox{P}\{Y(t)=1|X\} = [1+\exp\{-2-2t+ f_Y(X)\}]^{-1}$. The more complex nonlinear outcome model, for $t=0,1$, is specified as $Y(t) = 2+4.4t + f_t(X)+ \varepsilon_t$, where $f_t(X)= (7-4t)X_1 -\{(6+3t)X_6\}\{0.5+(X_2 + 1.4)^{(2+2t)}\}^{-1} + 2X_5^2 + 4X_8$, and  $\varepsilon_t\sim \mbox{N}(0, 1)$. In order to uphold the assumption of unconfoundedness, a selected subset has to include one of the subsets, 
$\{X_1,X_2,X_7\}$ or $\{X_1,X_2,X_8\}$.

\subsubsection{Setting 2: $M$-bias given $X$}\label{Sec:512}
In this setting the causal structure corresponds to Figure \ref{Fig:2}. 
Here $X_4$, $X_9$, $T$, $Y(0)$ and $Y(1)$ all depend on the unobservable variables $U_1$, $U_2$ and $U_3$. Everything else is analogous to the data generating process described in Section \ref{Setting1}. Now, let $U_1, \,U_2, \, U_3 \sim \mbox{N}(0,1),\,\,\nu_1, \,\nu_2 \sim \mbox{N}(0,0.5),\,\, X_4=0.2+0.8U_3+\nu_1$, and $X_9=1+2U_1+3U_2+\nu_2$. Here the outcome model functions are 
$f_T(X)=3 -2X_1 -2X_2 -2X_3 -X_4 -2X_7-U_1$,  $f_Y(X)=4X_1+ 2X_2 + 2X_5+ 4X_6 + 4X_8+7U_2+2U_3$, and, for $t=0, 1$, $f_t(X)= (7-3t)X_1 -\{(6+3t)X_6\}\{0.5+(X_2 + 1.4)^{(2+2t)}\}^{-1} + 2X_5^2 + 4X_8+7U_2+2U_3$. With these alterations the treatment variable $T$ and outcomes for the linear, binary and nonlinear models are generated following the specification in Section \ref{Setting1}.

\subsection{Estimation of the ACE}\label{Sec:52}
To illustrate the impact of the estimated target subsets on the estimation of the ACE we consider two different strategies: propensity score matching  \citep[PSM;][]{AI:06} and targeted maximum likelihood estimation \citep[TMLE;][]{vdLR:06}. PSM has several downsides \citep[see, e.g.,][]{KN:16} but is possibly the most popular strategy in practice. TMLE is a doubly robust estimator, i.e., consistent if either propensity score or outcome model is correct, and consistent and efficient if both models are correct. For PSM the propensity score is estimated by main effects logistic regression. One-to-one matching with replacement, and Euclidean distance as matching criterion, is used. For TMLE both the propensity score and outcome model is estimated by Bayesian additive regression trees \citep[BART;][]{CGM:10}. BART is a nonparametric regression method that have been shown to perform well in finite samples \citep{JH:11}.  
                 
   \subsubsection{Implementation details}
MMPC and MMHC are computed using the functions \texttt{mmpc} and \texttt{mmhc} in the package \texttt{bnlearn} \citep{MS:10}. The argument \texttt{optimized} is set to \texttt{FALSE} and for MMHC \texttt{score="aic"}. RF is computed using the function \texttt{randomForest} in the package \texttt{randomForest} \citep{LW:02}. Variables with \texttt{importance} larger than 25\% of the largest \texttt{importance} are included in the estimated set. LASSO is computed using the function \texttt{cv.glmnet} in the package \texttt{glmnet} \citep*{FHT:10} and variables with nonzero coefficients at \texttt{lambda.1se} are included in the estimated set. The LASSO model is specified to always include main effects, quadratic terms for the continuous covariates and all two-way interactions. $T$ and the discrete covariates are treated as factors for all four methods. For MMPC and MMHC continuous covariates and $Y$ are first discretized (using \texttt{discretize} with \texttt{method="quantile"} in \texttt{bnlearn}) and subsequently treated as factors. For RF, LASSO and when estimating the propensity score, continuous variables are not discretized. If $\widehat{ S}=\emptyset$, where $\widehat{S}$ is an estimate of the target covariate subset $S$, the propensity score is estimated as the proportion of treated units. The ACE estimators are evaluated with $\widehat{S}$ equal to each of the following covariate sets: $X$, $ \widehat{X}_{\rightarrow T}$, $\widehat{Q}_{\rightarrow T}$, $\widehat{X}_{\rightarrow Y} $, $\widehat{Z}_{\rightarrow Y} $, $\widehat{X}_{\rightarrow T, Y} $ and  $\widehat{W}_{\rightarrow Y}$. PSM is performed using the function \texttt{Match} in the package \texttt{Matching} \citep{JS:11}. TMLE estimates are computed using the functions \texttt{bartMachine} in package \texttt{bartMachine} \citep{KB:16} and \texttt{tmle} in package \texttt{tmle} \citep{GvdL:12}, default argument values are used.

\subsection{Simulation Results}\label{Res1}
\begin{figure}
\centerline{%
\includegraphics[width=19cm]{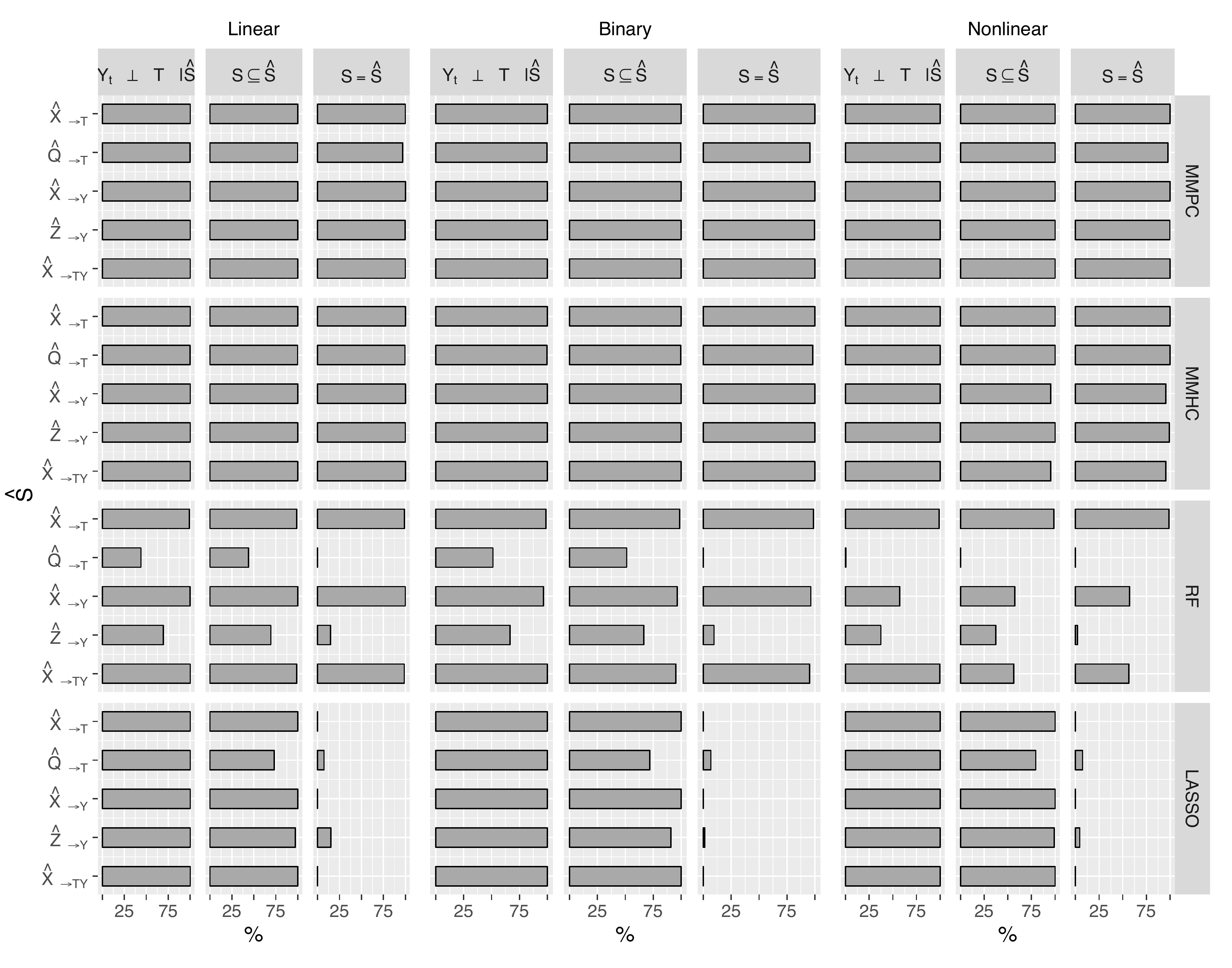}
}
\caption{Simulation results for Setting 1, $n=2000$.  Selection success rates (\%) for the covariate sets ($\hat{S}$) selected by MMPC, MMHC, RF or LASSO. Definitions of success are; i) unconfoundedness holds ($Y_t \perp T| \hat{S}$), ii) the target subset is included in the selected subset ($S  \subseteq \hat{S}$) and iii) equal subsets ($S  = \hat{S}$). The selected covariates sets are: covariates predicting treatment $\widehat{X}_{\rightarrow T}$, covariates predicting outcome $\widehat{X}_{\rightarrow Y}$, covariates predicting both treatment and outcome $\widehat{Q}_{\rightarrow T}\subseteq\widehat{X}_{\rightarrow T}$, $\widehat{Z}_{\rightarrow Y}\subseteq\widehat{X}_{\rightarrow Y}$ and $\widehat{X}_{\rightarrow T, Y}=\widehat{X}_{\rightarrow T}\cup \widehat{X}_{\rightarrow Y}$.}
\label{Fig:3}
\end{figure}

The results from the covariate selection algorithms are summarized in Tables 1-24 in Appendix B, where selection success rates and median cardinality of the selected sets are presented. Three definitions of success are used for the selected subset, $\widehat{S}$; i) unconfoundedness holds, i.e., $\big(Y(1),Y(0)\big) \perp\!\!\!\perp T \mid \widehat{S}$, ii) the target subset is included in the selected subset ($S  \subseteq \widehat{S}$), and iii) equal subsets ($S  = \widehat{S}$). The tables also include empirical bias, standard deviation and MSE for the ACE estimation as well as confidence interval coverage, mean width and mean lower and upper confidence interval limits. Results for Setting 1, $n=2000$, are illustrated in Figures \ref{Fig:3}-\ref{Fig:4}. Results for $\widehat{W}_{\rightarrow Y}$ are omitted throughout due to the fact that when the causal structure is estimated $\widehat{W}_{\rightarrow Y}$  and $\widehat{X}_{\rightarrow Y}$ turn out to be virtually identical.
 
 In Setting 1, when the sample size is relatively small ($n=500$, $n=1000$) neither of the network algorithms succeed in selecting only sets that uphold unconfoundedness. MMPC has in these cases higher rates of success than MMHC. For the linear outcome model, when $n=500$ the success rates for MMPC and MMHC are in the range [63.6, 99.2] and [31.6, 88.2], respectively, and for $n=1000$ the ranges are [94.9, 100.0] and [66.1, 98.8]. However, when the sample size is relatively large ($n=2000$, $n=10000$) virtually all selected sets have 100\% success rate in upholding unconfoundedness (the exception is the $\widehat{Q}_{\rightarrow T}$ sets with success rates 99.8 when $n=2000$), and this is the case for both network algorithms. For the larger sample sizes MMPC and MMHC not only select sets that uphold unconfoundedness, they frequently manage to exactly select the target subsets. For the binary outcome model, when $n=500$ the success rates for MMPC and MMHC are in the range [57.3, 97.7] and [29.0, 85.4], respectively, and for $n=1000$ the ranges are [92.8, 100.0] and [70.7, 99.2].  For the larger sample sizes the results are similar to the linear outcome case. For the nonlinear outcome model, when $n=500$ the success rates for MMPC and MMHC are in the range [83.2, 99.9] and [43.6, 99.7], respectively, and for $n=1000$ the success rates are 100.0 for all sets selected by MMPC and in the range [96.9, 100.0] for MMHC.  For the larger sample sizes all sets have 100\% success rate and both methods frequently manage to exactly select the target subsets.

RF performs similar to MMPC and MMHC with the important exception that it is, regardless of outcome model, unable to select the minimal target subsets $Q_{\rightarrow T}$ and $Z_{\rightarrow Y}$ with any desirable accuracy. Also, for the nonlinear outcome model case, RF fails when estimating the set $X_{\rightarrow Y}$.

Due to the fact that implementing the LASSO in a high-dimensional covariate space was much more time consuming than the other three methods, the investigation of LASSO is limited to $n=500, 1000, 2000$. LASSO is the only method that is able, regardless of sample size, to select sets that virtually always uphold unconfoundedness. However, sets selected by LASSO have much higher cardinality than the sets selected by the other three methods. Thus the dimension reduction is not as pronounced as with the other methods and the exact target sets are rarely selected. When $n=500, 1000$, LASSO performs better than the other methods, while for the larger sample sizes sets selected by any of the network methods often result in smaller MSE than sets selected by LASSO.

The simulation settings for the linear and binary outcome cases in this paper are similar to the linear and binary outcome cases in \cite{PHWdL:13} which allow us to make some comparisons of the methods used here and the kernel smoothing method used in the latter paper. Note that the kernel smoothing procedure is very computer intensive and is not feasible for the relatively high dimensions studied in this paper. For $n=1000$ (the largest sample size studied in \cite{PHWdL:13}) and only 10 covariates in $X$ kernel smoothing performs marginally better than MMPC does (when $n=1000$ and 100 covariates), in terms of success rates in upholding unconfoundedness, but MMPC manages to exactly retrieve the target subsets much more frequently. MMHC on the other hand is outperformed by kernel smoothing for such a relatively small sample size. 

When PSM is used, conditioning on one of the sets $X$, $ \widehat{X}_{\rightarrow T}$ or $\widehat{X}_{\rightarrow T, Y}$ results in a considerable larger bias and variance than conditioning on any of the other sets (where success rates of upholding unconfoundedness are 100\%). These results corroborates the results for PSM in \cite{PHWdL:13} in so far that, with very few exceptions, conditioning on $\widehat{Q}_{\rightarrow T}$ results in lower MSE than conditioning on $ \widehat{X}_{\rightarrow T}$ but that conditioning on $\widehat{X}_{\rightarrow Y} $ or $\widehat{Z}_{\rightarrow Y}$ in turn results in lower MSE than conditioning on $\widehat{Q}_{\rightarrow T}$. However, as pointed out by an anonymous referee, these results are probably, at least partly, driven by the fact that PSM with main effects logistic regression is a poor estimation strategy, and not by the selected sets per se. Consequently, when TMLE together with BART is used, first of all we see that MSEs generally are much lower than when using PSM with logistic regression. Secondly, we see that in many cases reducing the covariate set prior to estimating ACE results in higher MSE compared to using $X$. Still, for large sample sizes, conditioning on $\widehat{X}_{\rightarrow Y} $ results in a reduction in MSE compared to conditioning on $X$ and conditioning on $\widehat{X}_{\rightarrow T, Y}$ results in reduced or equal MSE compared to $X$.

\begin{figure}
\centerline{%
\includegraphics[width=19cm]{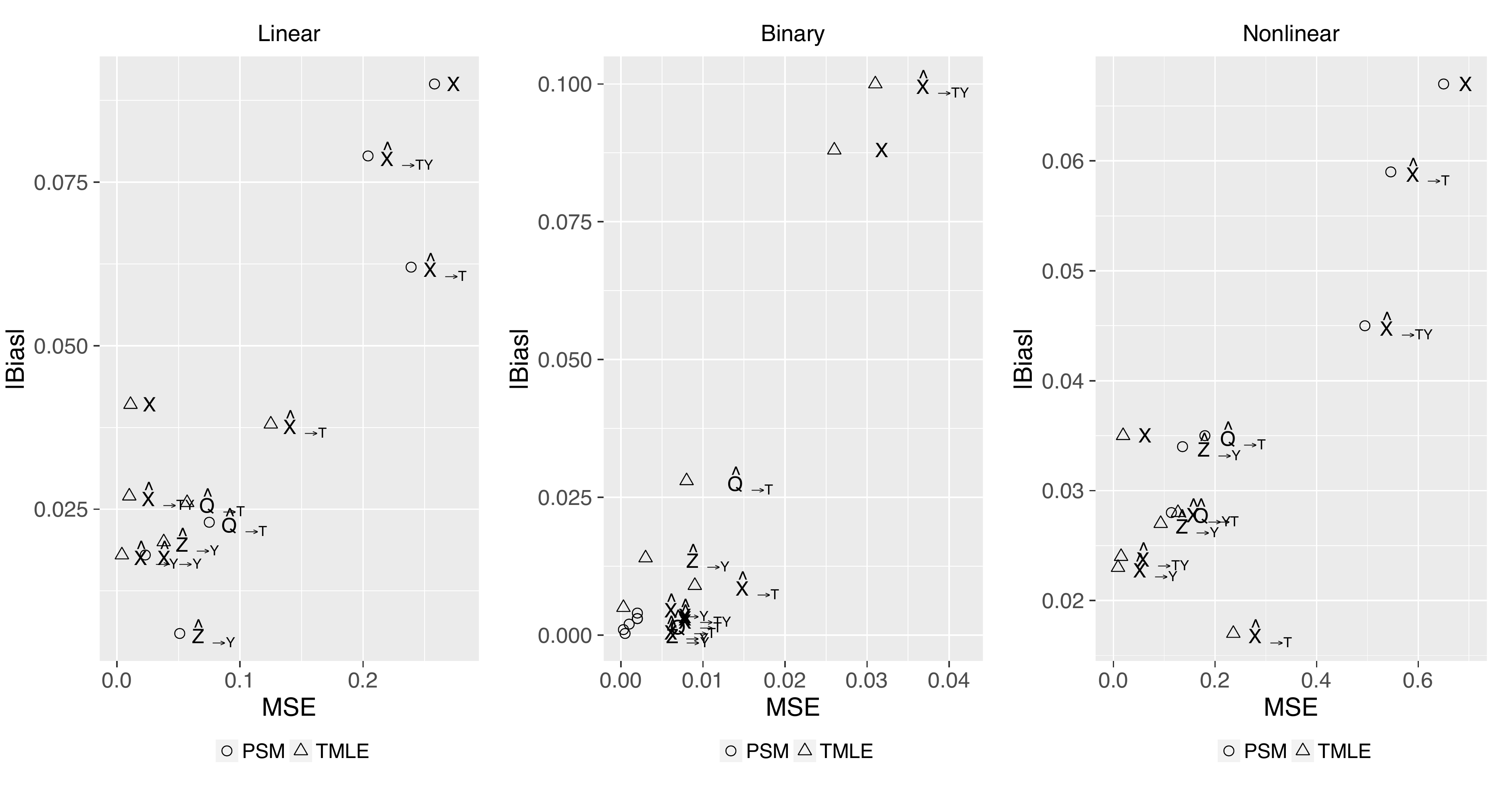}
}
\caption{Simulation results for Setting 1, $n=2000$. Absolute bias and MSE are for the ACE estimated using different sets of covariates as selected by MMPC and either propensity score matching (PSM) or targeted maximum likelihood estimation (TMLE). The different covariates sets are: $X$, covariates predicting treatment $\widehat{X}_{\rightarrow T}$, covariates predicting outcome $\widehat{X}_{\rightarrow Y}$, covariates predicting both treatment and outcome $\widehat{Q}_{\rightarrow T}\subseteq\widehat{X}_{\rightarrow T}$, $\widehat{Z}_{\rightarrow Y}\subseteq\widehat{X}_{\rightarrow Y}$ and $\widehat{X}_{\rightarrow T, Y}=\widehat{X}_{\rightarrow T}\cup \widehat{X}_{\rightarrow Y}$.}
\label{Fig:4}
\end{figure}

In Setting 2, when unconfoundedness does not hold given $X$ the success rates are, as expected, very low. In Setting 2 there are two ways in which a set can fail to uphold unconfoundedness: 1) if it does not include $X_4$ and/or 2) if it includes $X_9$. For the larger sample sizes all methods select sets that include both $X_4$ and $X_9$ or include too few covariates, thus failing the unconfoundedness assumption.  However, for PSM, except for $\widehat{X}_{\rightarrow T, Y}$, the sets selected by MMPC or MMHC often result in smaller bias and MSE compared to bias and MSE when conditioning on $X$. For TMLE it is only $\widehat{Q}_{\rightarrow T}$, $\widehat{X}_{\rightarrow Y}$ and $\widehat{Z}_{\rightarrow Y}$ that results in reduced MSE compared to $X$ for $n=10000$. Thus, in a situation where we are not perfectly sure that unconfoundedness is upheld when conditioning on $X$, at least for PSM, it does not seem to be harmful to use this confounder selection procedure for reducing the covariate set. 

As mentioned in Section \ref{Res1} implementing LASSO was more computer intensive than any of the methods MMPC, MMHC or RF. In Table \ref{Tab:comp} the computational times (measured by \texttt{system.time}) for $n=10000$ when only running the step where $ \widehat{X}_{\rightarrow T}$ is estimated from $X$, i.e., \texttt{mmpc}, \texttt{mmhc}, \texttt{cv.glmnet} and \texttt{randomForest} are each run only once. Timings differ slightly between runs but the example in Table \ref{Tab:comp} give an accurate description of the difference between methods. The computations were run on a MacBook Pro (Early 2015) with 3,1 GHz Intel Core i7 Processor and 16 GB 1867 MHz DDR3 Memory.

\begin{table}
 \caption{Computation time in seconds (and relative to MMPC) for $n = 10000$ when only running the step were $X_T$ is estimated from $\{X, T\}$, i.e., \texttt{mmpc}, \texttt{mmhc}, \texttt{cv.glmnet} and \texttt{randomForest} are each run only once.}
\label{Tab:comp}
\begin{center}
\begin{tabular}{lrr}
\hline \hline
Method & Seconds & Rel  \\
  \hline
MMPC&1.1 &  1.0\\
MMHC& 2.2&  2.0\\
RF&78.2 & 71.1\\
LASSO& 2114.3&1922.1  \\
\hline 
\end{tabular}
\end{center}
\end{table}

\section{Data Analysis}\label{Sec:6}
Previous studies have indicated that children delivered by C-section are at an increased risk of developing wheezing and asthma \citep[see, e.g., ][]{MM:11, LB:13}. Here, we select covariates for estimating the ACE of being delivered by C-section on, before the age of four, being prescribed medication commonly used to treat asthma. Using record linkage register data from the Umeå SIMSAM Lab \citep*{LNdLI:16} all children being the result of first-time mothers giving birth to full term (37 or more full weeks of gestation) singleton live offspring in the year 2006 in Sweden were identified ($n=41 857$). From this population the subset of children being delivered by C-section or non-instrumental vaginal delivery,  who were residents in Sweden during the whole study period, were offspring of two native Swedish parents and had no major malformations reported at birth were selected ($n=23 817$). Using data from the Swedish Prescribed Drug Registry all children being prescribed drugs with one of the ATC-codes \citep{ATC} R03AC, R03AK, R03BA, R03BC, R03CC, R03DC (hereinafter "asthma drugs") at least once before the age of four were identified.

As potential confounders we included 24 variables, listed in Table 25 in Appendix C. All observations with missing data on any of the covariates were excluded ($n=2950$), resulting in a complete cases sample containing $n=20867$ children. The proportion of children delivered by C-section was 20.9\% and 39.7\% of the children had been prescribed asthma drugs at least once before the age of four. MMPC and MMHC resulted in equal sets, namely:  $ \widehat{X}_{\rightarrow T}=$\{Maternal BMI at first antenatal visit, Gestational age\}, $ \widehat{Q}_{\rightarrow T}=$\{Maternal BMI at first antenatal visit\},  $ \widehat{X}_{\rightarrow Y}=$\{Maternal asthma, Paternal asthma drugs prescription within 6 months before delivery, Offspring sex, Birth place\},  $ \widehat{Z}_{\rightarrow Y}=$\{Offspring sex, Birth place\}, $ \widehat{X}_{\rightarrow T, Y} =\widehat{X}_{\rightarrow T} \cup \widehat{X}_{\rightarrow Y} $. The DAGs estimated by MMHC are given in Figures 1-7 in Appendix C (Cytoscape was used to visualize the graphs \citep{PS:03}). PSM and TMLE were implemented as described in Section \ref{Sec:52} and estimates of the ACE, i.e., risk difference, are presented in Table \ref{Tab:5}. For PSM the distributions of the estimated propensity scores for the different treatment groups were similar and exact matches were found when controlling for all sets except $X$ (where a caliper of 0.1 standard deviations of the estimated propensity score was used). The estimate based on raw data, not controlling for any confounders, suggest on average an 3.5\% risk increase in being prescribed asthma drugs before the age of four if you are delivered by C-section compared to what would be the risk in case of non-instrumental vaginal delivery. For both PSM and TMLE, controlling for any of the selected covariate sets results in slightly lower point estimates, although all still statistically significant. Assuming that we have unconfoundedness given the 24 potential confounders we started with, and taking into consideration the results from the simulation study where controlling for $ \widehat{X}_{\rightarrow Y}$ often was the best choice, the results in Table \ref{Tab:5} suggest that the ACE lies in the range [0.012, 0.047].

\begin{table}
 \caption{Estimates of the ACE. The cardinalities of the covariate sets (\#), ACE estimates ($\hat{\beta}$), standard errors (SE; Abadie-Imbens for PSM and influence-curved based for TMLE), lower (CIL) and upper (CIU) limits of 95\% confidence intervals. }
\label{Tab:5}
\begin{center}
\begin{tabular}{lccccc}
\hline \hline
 $\hat{S}$ & \# & $\hat{\beta}$ & SE & CIL & CIU \\  \hline
$\emptyset$ & 0 & 0.035 &0.008 &0.019 &0.052 \\
\hline
\multicolumn{6}{c}{PSM}\\
\hline
$X$ & 24&0.023  &0.010 & 0.004&0.042 \\
$\widehat{X}_{\rightarrow T}$ & 2&0.026  &0.009 &0.009 & 0.043\\
$\widehat{Q}_{\rightarrow T}$ &1 & 0.031 &0.008 &0.015 &0.048 \\
$  \widehat{X}_{\rightarrow Y}$ & 4& 0.029 &0.008 &0.013 &0.045 \\
$ \widehat{Z}_{\rightarrow Y}$ &2 & 0.031& 0.008&0.015 &0.048 \\
$ \widehat{X}_{\rightarrow T, Y} $ &6 & 0.021 &0.009 &0.004 &0.038 \\
\hline 
\multicolumn{6}{c}{TMLE}\\
\hline
$X$ & 24& 0.024 &0.011&0.003 &0.044 \\
$\widehat{X}_{\rightarrow T}$ & 2& 0.027 &0.009 &0.009 &0.045 \\
$\widehat{Q}_{\rightarrow T}$ &1 & 0.032 &0.009 &0.014 &0.049 \\
$  \widehat{X}_{\rightarrow Y}$ & 4& 0.029 &0.009&0.012&0.047 \\
$ \widehat{Z}_{\rightarrow Y}$ &2 &0.031& 0.009&0.014 &0.048 \\
$ \widehat{X}_{\rightarrow T, Y} $ &6 & 0.025& 0.010& 0.006&0.044 \\
\hline 
\end{tabular}
\end{center}
\end{table}

\section{Discussion}\label{Sec:7}

In this paper, we have introduced the network algorithms MMPC and MMHC in conjunction with the covariate selection algorithms in \cite{dLWR:11} as methods for confounder selection in causal inference when the true causal structure is not known. Given that unconfoundedness holds when conditioning on $X$,  the approach was shown, for sufficiently large sample sizes, to accurately estimate certain target covariate subsets. Compared to RF and LASSO, the network algorithms were preferable both with regard to estimation of the ACE and with regard to computational efficiency. However, it is very likely that the performance of RF and LASSO could be improved upon by carefully selecting their respective tuning parameter (variable importance cut-off and regularization parameter). Also, as expected, none of the four methods investigated were able to select covariate sets that uphold unconfoundedness when the true causal structure included a collider of unmeasured causes of the outcome and treatment. This is due to the fact that the methods cannot distinguish association from causation and thus a collider will frequently be included in the selected covariate set. How much one in practice should worry about the $M$-bias scenario exemplified in Setting \ref{Sec:512} is debatable.  \cite{DR:09},  \cite{WL:12}  and \cite{DM:15} suggest that it is rather uncommon and might be more of mathematical than practical interest. Moreover, the simulation results show that even if $M$-bias is present reducing the the dimension of the covariate set might still be beneficial.

The real data analysis consisted of 20867 observations and 24 covariates and this relatively high dimensional data proved to be more than feasible for MMPC and MMHC.




\section*{Acknowledgements}

This work was supported by the Swedish Research Council (Dnr: 2013-672). The simulations were performed on resources
provided by the Swedish National Infrastructure for Computing (SNIC) at High Performance Computing Centre North (HPC2N). The Umeå SIMSAM Lab data infrastructure used in this study was developed with support from the Swedish Research Council and by strategic funds from Umeå University. The author is grateful to the Co-Editor, the Associate Editor, and the referee for their helpful and constructive comments.\vspace*{-8pt}



%

\bibliographystyle{biom} \bibliography{Referenser.bib}

\begin{thebibliography}{}

\bibitem[\protect\citeauthoryear{Abadie and Imbens}{Abadie and
  Imbens}{2006}]{AI:06}
Abadie, A. and Imbens, G.~W. (2006).
\newblock Large sample properties of matching estimators for average treatment
  effects.
\newblock {\em Econometrica} {\bf 74,} 235--267.

\bibitem[\protect\citeauthoryear{Br{\aa}b{\"a}ck, Ek{\'e}us, Lowe, and
  Hjern}{Br{\aa}b{\"a}ck et~al.}{2013}]{LB:13}
Br{\aa}b{\"a}ck, L., Ek{\'e}us, C., Lowe, A.~J., and Hjern, A. (2013).
\newblock Confounding with familial determinants affects the association
  between mode of delivery and childhood asthma medication--a national cohort
  study.
\newblock {\em Allergy, Asthma \& Clinical Immunology} {\bf 9,} 1--9.

\bibitem[\protect\citeauthoryear{Breiman}{Breiman}{2001}]{LB:01}
Breiman, L. (2001).
\newblock Random forests.
\newblock {\em Machine Learning} {\bf 45,} 5--32.

\bibitem[\protect\citeauthoryear{Chickering}{Chickering}{2002}]{DC:02}
Chickering, D.~M. (2002).
\newblock Optimal structure identification with greedy search.
\newblock {\em Journal of Machine Learning Research} {\bf 3,} 507--554.

\bibitem[\protect\citeauthoryear{Chipman, George, and McCulloch}{Chipman
  et~al.}{2010}]{CGM:10}
Chipman, H.~A., George, E.~I., and McCulloch, R.~E. (2010).
\newblock {BART}: Bayesian additive regression trees.
\newblock {\em The Annals of Applied Statistics} {\bf 4,} 266--298.

\bibitem[\protect\citeauthoryear{Dawid}{Dawid}{1979}]{PD:79}
Dawid, A.~P. (1979).
\newblock Conditional independence in statistical theory.
\newblock {\em Journal of the Royal Statistical Society, Series B} {\bf 41,}
  1--31.

\bibitem[\protect\citeauthoryear{de~Luna, Waernbaum, and Richardson}{de~Luna
  et~al.}{2011}]{dLWR:11}
de~Luna, X., Waernbaum, I., and Richardson, T.~S. (2011).
\newblock Covariate selection for the nonparametric estimation of an average
  treatment effect.
\newblock {\em Biometrika} {\bf 98,} 861--875.

\bibitem[\protect\citeauthoryear{Ding and Miratrix}{Ding and
  Miratrix}{2015}]{DM:15}
Ding, P. and Miratrix, L.~W. (2015).
\newblock To adjust or not to adjust? {S}ensitivity analysis of {M}-bias and
  {B}utterfly-bias.
\newblock {\em Journal of Causal Inference} {\bf 3,} 41--57.

\bibitem[\protect\citeauthoryear{Friedman, Hastie, and Tibshirani}{Friedman
  et~al.}{2010}]{FHT:10}
Friedman, J., Hastie, T., and Tibshirani, R. (2010).
\newblock Regularization paths for generalized linear models via coordinate
  descent.
\newblock {\em Journal of Statistical Software} {\bf 33,} 1--22.

\bibitem[\protect\citeauthoryear{Friedman, Nachman, and Pe{\'e}r}{Friedman
  et~al.}{1999}]{FNP:99}
Friedman, N., Nachman, I., and Pe{\'e}r, D. (1999).
\newblock Learning {B}ayesian network structure from massive datasets: the
  "sparse candidate" algorithm.
\newblock In {\em Proceedings of the Fifteenth conference on Uncertainty in
  artificial intelligence}, pages 206--215. Morgan Kaufmann Publishers Inc.

\bibitem[\protect\citeauthoryear{Greenland}{Greenland}{2003}]{SG:03}
Greenland, S. (2003).
\newblock Quantifying biases in causal models: {C}lassical confounding vs
  collider-stratification bias.
\newblock {\em Epidemiology} {\bf 14,} 300--306.

\bibitem[\protect\citeauthoryear{Gruber and {van der Laan}}{Gruber and {van der
  Laan}}{2012}]{GvdL:12}
Gruber, S. and {van der Laan}, M.~J. (2012).
\newblock {tmle}: An {R} package for targeted maximum likelihood estimation.
\newblock {\em Journal of Statistical Software} {\bf 51,} 1--35.

\bibitem[\protect\citeauthoryear{Hill}{Hill}{2011}]{JH:11}
Hill, J.~L. (2011).
\newblock Bayesian nonparametric modeling for causal inference.
\newblock {\em Journal of Computational and Graphical Statistics} {\bf 20,}
  217--240.

\bibitem[\protect\citeauthoryear{Häggström}{Häggström}{2016}]{CovSelHigh}
Häggström, J. (2016).
\newblock {\em \textit{{C}ov{S}el{H}igh: {M}odel-{F}ree {C}ovariate {S}election
  in {H}igh {D}imensions}}.
\newblock R package version 1.0.0.

\bibitem[\protect\citeauthoryear{Häggström, Persson, Waernbaum, and
  de~Luna}{Häggström et~al.}{2015}]{CovSel}
Häggström, J., Persson, E., Waernbaum, I., and de~Luna, X. (2015).
\newblock Cov{S}el: An {R} package for covariate selection when estimating
  average causal effects.
\newblock {\em Journal of Statistical Software} {\bf 68,} 1--20.

\bibitem[\protect\citeauthoryear{Kapelner and Bleich}{Kapelner and
  Bleich}{2016}]{KB:16}
Kapelner, A. and Bleich, J. (2016).
\newblock {bartMachine}: Machine learning with {B}ayesian additive regression
  trees.
\newblock {\em Journal of Statistical Software} {\bf 70,} 1--40.

\bibitem[\protect\citeauthoryear{King and Nielsen}{King and
  Nielsen}{2016}]{KN:16}
King, G. and Nielsen, R. (2016).
\newblock Why propensity scores should not be used for matching.
\newblock Working paper. Copy at http://j.mp/1sexgVw.

\bibitem[\protect\citeauthoryear{Liaw and Wiener}{Liaw and
  Wiener}{2002}]{LW:02}
Liaw, A. and Wiener, M. (2002).
\newblock Classification and regression by random{F}orest.
\newblock {\em R News} {\bf 2,} 18--22.

\bibitem[\protect\citeauthoryear{Lindgren, Nilsson, de~Luna, and
  Ivarsson}{Lindgren et~al.}{2016}]{LNdLI:16}
Lindgren, U., Nilsson, K., de~Luna, X., and Ivarsson, A. (2016).
\newblock Data resource profile: {S}wedish microdata research from childhood
  into lifelong health and welfare ({U}me{\aa} {SIMSAM} {L}ab).
\newblock {\em International Journal of Epidemiology} {\bf 45,} 1075--1075g.

\bibitem[\protect\citeauthoryear{Liu, Brookhart, Schneeweiss, Mi, and
  Setoguchi}{Liu et~al.}{2012}]{WL:12}
Liu, W., Brookhart, M.~A., Schneeweiss, S., Mi, X., and Setoguchi, S. (2012).
\newblock Implications of {M} bias in epidemiologic studies: A simulation
  study.
\newblock {\em American Journal of Epidemiology} {\bf 176,} 938--948.

\bibitem[\protect\citeauthoryear{Maathuis, Kalisch, and Bühlmann}{Maathuis
  et~al.}{2009}]{MKB:09}
Maathuis, M.~H., Kalisch, M., and Bühlmann, P. (2009).
\newblock Estimating high-dimensional intervention effects from observational
  data.
\newblock {\em The Annals of Statistics} {\bf 37,} 3133--3164.

\bibitem[\protect\citeauthoryear{Magnus, H{\aa}berg, Stigum, Nafstad, London,
  Vangen, and Nystad}{Magnus et~al.}{2011}]{MM:11}
Magnus, M.~C., H{\aa}berg, S.~E., Stigum, H., Nafstad, P., London, S.~J.,
  Vangen, S., and Nystad, W. (2011).
\newblock Delivery by {C}esarean section and early childhood respiratory
  symptoms and disorders: the {N}orwegian mother and child cohort study.
\newblock {\em American Journal of Epidemiology} {\bf 174,} 1275--1285.

\bibitem[\protect\citeauthoryear{Neyman}{Neyman}{1923}]{JN:23}
Neyman, J. (1923).
\newblock On the application of probability theory to agricultural experiments,
  essay on principles. \emph{Roczniki nauk Rolczych X}, 1-51. {In Polish}.
\newblock {\em \emph{English translation by D.M. Dabrowska and T.P. Speed in}
  Statistical Science} {\bf 5,} 465--472.

\bibitem[\protect\citeauthoryear{Pearl}{Pearl}{1995}]{JP:95}
Pearl, J. (1995).
\newblock Causal diagrams for empirical research.
\newblock {\em Biometrika} {\bf 82,} 669--688.

\bibitem[\protect\citeauthoryear{Pearl}{Pearl}{2009a}]{JP:09}
Pearl, J. (2009a).
\newblock {\em Causality, Second edition}.
\newblock Cambridge University Press, Cambridge.

\bibitem[\protect\citeauthoryear{Pearl}{Pearl}{2009b}]{JP:09a}
Pearl, J. (2009b).
\newblock Letter to the editor.
\newblock {\em Statistics in Medicine} {\bf 28,} 1415--1416.

\bibitem[\protect\citeauthoryear{{Persson}, {H{\"a}ggstr{\"o}m}, {Waernbaum},
  and {de Luna}}{{Persson} et~al.}{2017}]{PHWdL:13}
{Persson}, E., {H{\"a}ggstr{\"o}m}, J., {Waernbaum}, I., and {de Luna}, X.
  (2017).
\newblock Data-driven algorithms for dimension reduction in causal inference.
\newblock {\em Computational Statistics \& Data Analysis} {\bf 105,} 280--292.

\bibitem[\protect\citeauthoryear{{R Core Team}}{{R Core Team}}{2016}]{R}
{R Core Team} (2016).
\newblock {\em R: A Language and Environment for Statistical Computing}.
\newblock R Foundation for Statistical Computing, Vienna, Austria.

\bibitem[\protect\citeauthoryear{Rubin}{Rubin}{1974}]{DN:74}
Rubin, D.~B. (1974).
\newblock Estimating causal effects of treatments in randomized and
  nonrandomized studies.
\newblock {\em Journal of Educational Psychology} {\bf 66,} 688--701.

\bibitem[\protect\citeauthoryear{Rubin}{Rubin}{1990}]{DR:90}
Rubin, D.~B. (1990).
\newblock Formal modes of statistical inference for causal effects.
\newblock {\em Journal of Statistical Planning and Inference} {\bf 25,}
  279--292.

\bibitem[\protect\citeauthoryear{Rubin}{Rubin}{2007}]{DR:07}
Rubin, D.~B. (2007).
\newblock The design versus the analysis of observational studies for causal
  effects: Parallels with the design of randomized trials.
\newblock {\em Statistics in Medicine} {\bf 26,} 20--36.

\bibitem[\protect\citeauthoryear{Rubin}{Rubin}{2008a}]{DR:08}
Rubin, D.~B. (2008a).
\newblock Author's reply (to {I}an {S}hrier's letter to the editor).
\newblock {\em Statistics in Medicine} {\bf 27,} 2741--2742.

\bibitem[\protect\citeauthoryear{Rubin}{Rubin}{2008b}]{DR:08b}
Rubin, D.~B. (2008b).
\newblock For objective causal inference, design trumps analysis.
\newblock {\em The Annals of Applied Statistics} {\bf 2,} 808--840.

\bibitem[\protect\citeauthoryear{Rubin}{Rubin}{2009}]{DR:09}
Rubin, D.~B. (2009).
\newblock Author's reply (to {J}udea {P}earl's and {A}rvid {S}j\"olander's
  letters to the editor).
\newblock {\em Statistics in Medicine} {\bf 28,} 1420--1423.

\bibitem[\protect\citeauthoryear{Schnitzer, Lok, and Gruber}{Schnitzer
  et~al.}{2016}]{SLG:15}
Schnitzer, M.~E., Lok, J.~J., and Gruber, S. (2016).
\newblock Variable selection for confounder control, flexible modeling and
  collaborative targeted minimum loss-based estimation in causal inference.
\newblock {\em The International Journal of Biostatistics} {\bf 12,} 97--115.

\bibitem[\protect\citeauthoryear{Scutari}{Scutari}{2010}]{MS:10}
Scutari, M. (2010).
\newblock Learning {B}ayesian networks with the bnlearn {R} package.
\newblock {\em Journal of Statistical Software} {\bf 35,} 1--22.

\bibitem[\protect\citeauthoryear{Sekhon}{Sekhon}{2011}]{JS:11}
Sekhon, J.~S. (2011).
\newblock Multivariate and propensity score matching software with automated
  balance optimization: the {M}atching package for {R}.
\newblock {\em Journal of Statistical Software} {\bf 42,} 1--52.

\bibitem[\protect\citeauthoryear{Shannon, Markiel, Ozier, Baliga, Wang, Ramage,
  Amin, Schwikowski, and Ideker}{Shannon et~al.}{2003}]{PS:03}
Shannon, P., Markiel, A., Ozier, O., Baliga, N.~S., Wang, J.~T., Ramage, D.,
  Amin, N., Schwikowski, B., and Ideker, T. (2003).
\newblock Cytoscape: A software environment for integrated models of
  biomolecular interaction networks.
\newblock {\em Genome Research} {\bf 13,} 2498--2504.

\bibitem[\protect\citeauthoryear{Shrier}{Shrier}{2008}]{IS:08}
Shrier, I. (2008).
\newblock Letter to the editor.
\newblock {\em Statistics in Medicine} {\bf 27,} 2740--2741.

\bibitem[\protect\citeauthoryear{Sj{\"o}lander}{Sj{\"o}lander}{2009}]{AS:09}
Sj{\"o}lander, A. (2009).
\newblock Letter to the editor.
\newblock {\em Statistics in Medicine} {\bf 28,} 1416--1420.

\bibitem[\protect\citeauthoryear{Spirtes, Glymour, and Scheines}{Spirtes
  et~al.}{2000}]{SGS:00}
Spirtes, P., Glymour, C.~N., and Scheines, R. (2000).
\newblock {\em Causation, Prediction, and Search}.
\newblock MIT press.

\bibitem[\protect\citeauthoryear{Tibshirani}{Tibshirani}{1996}]{RT:86}
Tibshirani, R. (1996).
\newblock Regression shrinkage and selection via the lasso.
\newblock {\em Journal of the Royal Statistical Society. Series B
  (Methodological)} {\bf 58,} 267--288.

\bibitem[\protect\citeauthoryear{Tsamardinos, Brown, and Aliferis}{Tsamardinos
  et~al.}{2006}]{TI:06}
Tsamardinos, I., Brown, L., and Aliferis, C. (2006).
\newblock The max-min hill-climbing {B}ayesian network structure learning
  algorithm.
\newblock {\em Machine Learning} {\bf 65,} 31--78.

\bibitem[\protect\citeauthoryear{van~der Laan and Rubin}{van~der Laan and
  Rubin}{2006}]{vdLR:06}
van~der Laan, M.~J. and Rubin, D. (2006).
\newblock {T}argeted maximum likelihood learning.
\newblock {\em The International Journal of Biostatistics} {\bf 2,} 1--38.

\bibitem[\protect\citeauthoryear{VanderWeele and Shpitser}{VanderWeele and
  Shpitser}{2011}]{VS:11}
VanderWeele, T.~J. and Shpitser, I. (2011).
\newblock A new criterion for confounder selection.
\newblock {\em Biometrics} {\bf 67,} 1406--1413.

\bibitem[\protect\citeauthoryear{{WHO Collaborating Centre for Drug Statistics
  Methodology}}{{WHO Collaborating Centre for Drug Statistics
  Methodology}}{2014}]{ATC}
{WHO Collaborating Centre for Drug Statistics Methodology} (2014).
\newblock {\em {ATC} Classification Index With {DDD}s, 2015}.
\newblock WHO Collaborating Centre for Drug Statistics Methodology, Oslo,
  Norway.

\end{thebibliography}

 \section*{Appendix A: Proof of Theorem 1 in Section 4.4}
 \label{App:1}

\textit{Proof}. Let $G_{A,B}$ denote a graph involving only the variables in the sets $A$ and $B$. Consider the graphs $G_{X, T}$, $G_{X_ {\rightarrow T}, Y(t)}$, $G_{X, Y(t)}$, $G_{X_ {\rightarrow Y}, T}$ and $G_{X_ {\rightarrow T, Y}, Y(t)}$, $t=0,1$. 

It follows from C3 that the above graphs are all DAGs since each of them is a subgraph of a DAG.
 
It follows from C4-C5 that there exist joint probability distributions $p_{X, T}$, $p_{X_ {\rightarrow T}, Y(t)}$, $p_{X, Y(t)}$, $p_{X_ {\rightarrow Y}, T}$ and $p_{X_ {\rightarrow T, Y}, Y(t)}$, $t=0,1$, such that the local Markov property holds and which are faithful to the respective subgraph. 

Then, together with conditions C1-C2 and C6, it follows from Theorem 3 in \cite{TI:06} that if in the estimated skeleton of $G_{X, T}$, produced by MMPC with input variables $\{X, T\}$, there is an edge connecting $T$ to $X_j\in X$ then $X_j$ is a parent of $T$. Hence, $\widehat{X}_ {\rightarrow T}=X_ {\rightarrow T}$. Similarly, if in the estimated skeleton of $G_{X_ {\rightarrow T}, Y(t)}$, $t=0,1$, produced by MMPC with input variables $\{X_ {\rightarrow T}, Y|T=t\}$, there is an edge connecting $Y|T=t$ to $X_j\in X_ {\rightarrow T}$ then $X_j$ is a parent of $Y|T=t$, i.e., a parent of $Y(t)$ since $Y(t)\ci T|X$. Hence, $\widehat{Q}_ {\rightarrow T}^t=Q_ {\rightarrow T}^t$, $t=0,1$ and therefore $\widehat{Q}_ {\rightarrow T}=Q_ {\rightarrow T}$. Analogously, $\widehat{X}_{\rightarrow Y}=X_{\rightarrow Y}$ and $\widehat{Z}_{\rightarrow }=Z_{\rightarrow Y}$. Since $\widehat{X}_{\rightarrow T, Y}=\widehat{X}_{\rightarrow T}\cup \widehat{X}_{\rightarrow Y}$ it follows that if $\widehat{X}_ {\rightarrow T}=X_ {\rightarrow T}$ and $\widehat{X}_{\rightarrow Y}=X_{\rightarrow Y}$ then $\widehat{X}_{\rightarrow T, Y}=X_{\rightarrow T, Y}$. Finally, according to the same reasoning as above using MMPC with input variables $(X_ {\rightarrow T, Y}, Y|T=t)$ results in $\widehat{W}_{\rightarrow Y}=W_{\rightarrow Y}$

\section*{Appendix B: Simulation results from Section 5.3}
\label{App:2}

\begin{table}
\centering
 \def\~{\hphantom{0}}
 \caption{Setting 1, linear outcome model, $n=500, 1000$. The proportion (\%) of times the selected subset, $\widehat{S}$, satisfies three conditions of unconfoundedness and the median cardinality of $\widehat{S}$ (\#). Average bias (Bias), standard deviation (SD) and mean square error (MSE) from estimating ACE with PSM and TMLE. }
 \label{Tab:1a}
\begin{tabular}{lllrrrrrrrrrr}
\hline
\hline
\rule{0pt}{4ex}  
& \multicolumn{6}{c}{} & \multicolumn{3}{c}{PSM} &  \multicolumn{3}{c}{TMLE}  \\ 
   \cline{8-10}  \cline{11-13}
  $n$& \multicolumn{1}{c}{Method} & \multicolumn{1}{c}{$\widehat{S}$} &  \multicolumn{1}{c}{$Y_t \perp\!\!\!\perp T \mid \widehat{S}$} & \multicolumn{1}{c}{$S  \subseteq \widehat{S}$} & \multicolumn{1}{c}{$S=\widehat{S}$} &\multicolumn{1}{c}{ \#} & \multicolumn{1}{c}{Bias} & \multicolumn{1}{c}{SD} & \multicolumn{1}{c}{MSE} & \multicolumn{1}{c}{Bias} & \multicolumn{1}{c}{SD} & \multicolumn{1}{c}{MSE} \\ 
  \hline
500 &   & $X$ & 100.0 & 100.0 & 100.0 & 100 & 0.340 & 1.370 & 1.992 & 0.133 & 0.163 & 0.044 \\ 
  \cline{2-13}
  & MMPC & $\widehat{X}_{\rightarrow T}$ & 88.1 & 83.7 & 83.7 & 5 & 0.251 & 0.937 & 0.941 & 0.208 & 0.755 & 0.612 \\ 
    &   & $\widehat{Q}_{\rightarrow T}$ & 63.6 & 63.6 & 61.8 & 3 & 0.406 & 0.757 & 0.737 & 0.404 & 0.703 & 0.658 \\ 
    &   & $ \widehat{X}_{\rightarrow Y}$ & 88.5 & 27.3 & 27.3 & 4 & 0.195 & 0.501 & 0.289 & 0.189 & 0.392 & 0.189 \\ 
    &   & $\widehat{Z}_{\rightarrow Y}$ & 84.9 & 84.9 & 84.0 & 3 & 0.195 & 0.586 & 0.381 & 0.197 & 0.526 & 0.316 \\ 
    &   & $\widehat{X}_{\rightarrow T, Y}$ & 99.2 & 28.4 & 28.4 & 7 & 0.176 & 0.791 & 0.656 & 0.101 & 0.319 & 0.112 \\
\hline  
         \end{tabular}
\vspace{6pt}
 \end{table}

 \clearpage

\begin{table}
\centering
 \def\~{\hphantom{0}}
 \caption{Setting 1, linear outcome model, $n=2000, 10000$. The proportion (\%) of times the selected subset, $\widehat{S}$, satisfies three conditions of unconfoundedness and the median cardinality of $\widehat{S}$ (\#). Average bias (Bias), standard deviation (SD) and mean square error (MSE) from estimating ACE with PSM and TMLE. }
 \label{Tab:1b}
 \begin{tabular}{lllrrrrrrrrrr}
\hline
\hline
\rule{0pt}{4ex} 
 & \multicolumn{6}{c}{} & \multicolumn{3}{c}{PSM} &  \multicolumn{3}{c}{TMLE}  \\ 
   \cline{8-10}  \cline{11-13}
  $n$& \multicolumn{1}{c}{Method} & \multicolumn{1}{c}{$\widehat{S}$} &  \multicolumn{1}{c}{$Y_t \perp\!\!\!\perp T \mid \widehat{S}$} & \multicolumn{1}{c}{$S  \subseteq \widehat{S}$} & \multicolumn{1}{c}{$S=\widehat{S}$} &\multicolumn{1}{c}{ \#} & \multicolumn{1}{c}{Bias} & \multicolumn{1}{c}{SD} & \multicolumn{1}{c}{MSE} & \multicolumn{1}{c}{Bias} & \multicolumn{1}{c}{SD} & \multicolumn{1}{c}{MSE} \\ 
  \hline
 2000 &   &  $X$ & 100.0 & 100.0 & 100.0 & 100 & 0.090 & 0.500 & 0.258 & 0.041 & 0.099 & 0.011 \\ 
   \cline{2-13}
  & MMPC & $\widehat{X}_{\rightarrow T}$ & 100.0 & 100.0 & 99.9 & 5 & 0.062 & 0.484 & 0.239 & 0.038 & 0.351 & 0.125 \\ 
    &   & $\widehat{Q}_{\rightarrow T}$ & 99.8 & 99.8 & 96.8 & 3 & 0.023 & 0.274 & 0.075 & 0.026 & 0.237 & 0.057 \\ 
    &   & $ \widehat{X}_{\rightarrow Y}$ & 100.0 & 100.0 & 100.0 & 5 & 0.018 & 0.151 & 0.023 & 0.018 & 0.062 & 0.004 \\ 
    &   & $\widehat{Z}_{\rightarrow Y}$ & 100.0 & 100.0 & 99.8 & 3 & 0.006 & 0.227 & 0.051 & 0.020 & 0.193 & 0.038 \\ 
    &   & $\widehat{X}_{\rightarrow T, Y}$ & 100.0 & 100.0 & 99.9 & 8 & 0.079 & 0.444 & 0.204 & 0.027 & 0.097 & 0.010 \\
\hline
       \end{tabular}
\vspace{6pt}
 \end{table}

  \clearpage
 
\begin{table}
\centering
 \def\~{\hphantom{0}}
 \begin{minipage}{175mm}
 \caption{Setting 1, binary outcome model, $n=500, 1000$. The proportion (\%) of times the selected subset, $\widehat{S}$, satisfies three conditions of unconfoundedness and the median cardinality of $\widehat{S}$ (\#). Average bias (Bias), standard deviation (SD) and mean square error (MSE) from estimating ACE with PSM and TMLE.  }
 \label{Tab:1e}
 \begin{tabular}{lllrrrrrrrrrr}
\hline
\hline
\rule{0pt}{4ex}  
 & \multicolumn{6}{c}{} & \multicolumn{3}{c}{PSM} &  \multicolumn{3}{c}{TMLE}  \\ 
   \cline{8-10}  \cline{11-13}
  $n$& \multicolumn{1}{c}{Method} & \multicolumn{1}{c}{$\widehat{S}$} &  \multicolumn{1}{c}{$Y_t \perp\!\!\!\perp T \mid \widehat{S}$} & \multicolumn{1}{c}{$S  \subseteq \widehat{S}$} & \multicolumn{1}{c}{$S=\widehat{S}$} &\multicolumn{1}{c}{ \#} & \multicolumn{1}{c}{Bias} & \multicolumn{1}{c}{SD} & \multicolumn{1}{c}{MSE} & \multicolumn{1}{c}{Bias} & \multicolumn{1}{c}{SD} & \multicolumn{1}{c}{MSE} \\ 
  \hline
  500 &  &  $X$ & 100.0 & 100.0 & 100.0 & 100 & -0.016 & 0.132 & 0.018 & -0.066 & 0.079 & 0.011 \\ 
    \cline{2-13}
  & MMPC & $\widehat{X}_{\rightarrow T}$ & 88.1 & 83.7 & 83.7 & 5 & -0.011 & 0.081 & 0.007 & -0.037 & 0.147 & 0.023 \\ 
    &   & $\widehat{Q}_{\rightarrow T}$ & 57.3 & 57.3 & 55.6 & 3 & -0.025 & 0.062 & 0.004 & -0.074 & 0.121 & 0.020 \\ 
    &   & $ \widehat{X}_{\rightarrow Y}$ & 68.4 & 28.5 & 28.2 & 4 & -0.024 & 0.052 & 0.003 & -0.034 & 0.061 & 0.005 \\ 
    &   & $\widehat{Z}_{\rightarrow Y}$ & 66.4 & 66.4 & 65.9 & 3 & -0.021 & 0.056 & 0.004 & -0.048 & 0.095 & 0.011 \\ 
    &   & $\widehat{X}_{\rightarrow T, Y}$ & 97.7 & 38.1 & 37.8 & 7 & -0.009 & 0.077 & 0.006 & -0.056 & 0.147 & 0.025 \\
\hline    
     \end{tabular}
 \end{minipage}
\vspace{6pt}
 \end{table}
 
 \clearpage
 
\begin{table}
\centering
 \def\~{\hphantom{0}}
 \begin{minipage}{175mm}
 \caption{Setting 1, binary outcome model, $n=2000, 10000$. The proportion (\%) of times the selected subset, $\widehat{S}$, satisfies three conditions of unconfoundedness and the median cardinality of $\widehat{S}$ (\#). Average bias (Bias), standard deviation (SD) and mean square error (MSE) from estimating ACE with PSM and TMLE. }
 \label{Tab:1e}
 \begin{tabular}{lllrrrrrrrrrr}
\hline
\hline
\rule{0pt}{4ex}  
 & \multicolumn{6}{c}{} & \multicolumn{3}{c}{PSM} &  \multicolumn{3}{c}{TMLE}  \\ 
   \cline{8-10}  \cline{11-13}
  $n$& \multicolumn{1}{c}{Method} & \multicolumn{1}{c}{$\widehat{S}$} &  \multicolumn{1}{c}{$Y_t \perp\!\!\!\perp T \mid \widehat{S}$} & \multicolumn{1}{c}{$S  \subseteq \widehat{S}$} & \multicolumn{1}{c}{$S=\widehat{S}$} &\multicolumn{1}{c}{ \#} & \multicolumn{1}{c}{Bias} & \multicolumn{1}{c}{SD} & \multicolumn{1}{c}{MSE} & \multicolumn{1}{c}{Bias} & \multicolumn{1}{c}{SD} & \multicolumn{1}{c}{MSE} \\ 
  \hline
  2000& & $X$ & 100.0 & 100.0 & 100.0 & 100 & -0.003 & 0.046 & 0.002 & -0.088 & 0.137 & 0.026 \\ 
  \cline{2-13}
    & MMPC & $\widehat{X}_{\rightarrow T}$ & 100.0 & 100.0 & 99.9 & 5 & -0.003 & 0.043 & 0.002 & -0.009 & 0.094 & 0.009 \\ 
    &   & $\widehat{Q}_{\rightarrow T}$ & 99.6 & 99.6 & 95.7 & 3 & -0.002 & 0.026 & 0.001 & -0.028 & 0.083 & 0.008 \\ 
    &   & $ \widehat{X}_{\rightarrow Y}$ & 100.0 & 100.0 & 100.0 & 5 & -0.001 & 0.019 & 3e-04 & -0.005 & 0.016 & 3e-04 \\ 
    &   & $\widehat{Z}_{\rightarrow Y}$ & 100.0 & 100.0 & 99.8 & 3 & 2e-04 & 0.022 & 5e-04 & -0.014 & 0.056 & 0.003 \\ 
    &   & $\widehat{X}_{\rightarrow T, Y}$ & 100.0 & 100.0 & 99.9 & 8 & -0.004 & 0.041 & 0.002 & -0.100 & 0.145 & 0.031 \\
\hline       
     \end{tabular}
 \end{minipage}
\vspace{6pt}
 \end{table}
 
 \clearpage
  
\begin{table}
\centering
 \def\~{\hphantom{0}}
 \begin{minipage}{175mm}
 \caption{Setting 1, nonlinear outcome model, $n=500, 1000$. The proportion (\%) of times the selected subset, $\widehat{S}$, satisfies three conditions of unconfoundedness and the median cardinality of $\widehat{S}$ (\#). Average bias (Bias), standard deviation (SD) and mean square error (MSE) from estimating ACE with PSM and TMLE. }
 \label{Tab:1c}
 \begin{tabular}{lllrrrrrrrrrr}
\hline
\hline
\rule{0pt}{4ex}  
 & \multicolumn{6}{c}{} & \multicolumn{3}{c}{PSM} &  \multicolumn{3}{c}{TMLE}  \\ 
   \cline{8-10}  \cline{11-13}
  $n$& \multicolumn{1}{c}{Method} & \multicolumn{1}{c}{$\widehat{S}$} &  \multicolumn{1}{c}{$Y_t \perp\!\!\!\perp T \mid \widehat{S}$} & \multicolumn{1}{c}{$S  \subseteq \widehat{S}$} & \multicolumn{1}{c}{$S=\widehat{S}$} &\multicolumn{1}{c}{ \#} & \multicolumn{1}{c}{Bias} & \multicolumn{1}{c}{SD} & \multicolumn{1}{c}{MSE} & \multicolumn{1}{c}{Bias} & \multicolumn{1}{c}{SD} & \multicolumn{1}{c}{MSE} \\ 
  \hline    
500&   & $X$ & 100.0 & 100.0 & 100.0 & 100 & 0.286 & 2.077 & 4.396 & 0.013 & 0.279 & 0.078 \\
     \cline{2-13}
  & MMPC & $\widehat{X}_{\rightarrow T}$ & 88.1 & 83.7 & 83.7 & 5 & 0.275 & 1.310 & 1.791 & 0.086 & 0.998 & 1.003 \\ 
    &   & $\widehat{Q}_{\rightarrow T}$ & 83.2 & 83.2 & 81.8 & 3 & 0.172 & 0.906 & 0.850 & 0.145 & 0.791 & 0.647 \\ 
    &   & $ \widehat{X}_{\rightarrow Y}$ & 99.0 & 12.2 & 12.1 & 4 & -0.033 & 0.712 & 0.508 & -0.019 & 0.388 & 0.151 \\ 
    &   & $\widehat{Z}_{\rightarrow Y}$ & 95.2 & 95.2 & 94.4 & 3 & 0.022 & 0.756 & 0.572 & 0.010 & 0.639 & 0.408 \\ 
    &   & $\widehat{X}_{\rightarrow T, Y}$ & 99.9 & 11.2 & 11.1 & 7 & 0.154 & 1.198 & 1.459 & -0.051 & 0.513 & 0.266 \\
\hline 
    \end{tabular}
 \end{minipage}
\vspace{6pt}
 \end{table}
 
 \clearpage

\begin{table}
\centering
 \def\~{\hphantom{0}}
 \begin{minipage}{175mm}
 \caption{Setting 1, nonlinear outcome model, $n=2000, 10000$. The proportion (\%) of times the selected subset, $\widehat{S}$, satisfies three conditions of unconfoundedness and the median cardinality of $\widehat{S}$ (\#). Average bias (Bias), standard deviation (SD) and mean square error (MSE) from estimating ACE with PSM and TMLE.  }
 \label{Tab:1d}
 \begin{tabular}{lllrrrrrrrrrr}
\hline
\hline
\rule{0pt}{4ex} 
 & \multicolumn{6}{c}{} & \multicolumn{3}{c}{PSM} &  \multicolumn{3}{c}{TMLE}  \\ 
   \cline{8-10}  \cline{11-13}
  $n$& \multicolumn{1}{c}{Method} & \multicolumn{1}{c}{$\widehat{S}$} &  \multicolumn{1}{c}{$Y_t \perp\!\!\!\perp T \mid \widehat{S}$} & \multicolumn{1}{c}{$S  \subseteq \widehat{S}$} & \multicolumn{1}{c}{$S=\widehat{S}$} &\multicolumn{1}{c}{ \#} & \multicolumn{1}{c}{Bias} & \multicolumn{1}{c}{SD} & \multicolumn{1}{c}{MSE} & \multicolumn{1}{c}{Bias} & \multicolumn{1}{c}{SD} & \multicolumn{1}{c}{MSE} \\ 
  \hline    
 2000 &   & $X$ & 100.0 & 100.0 & 100.0 & 100 & 0.067 & 0.803 & 0.650 & -0.035 & 0.134 & 0.019 \\ 
\cline{2-13}
  & MMPC & $\widehat{X}_{\rightarrow T}$ & 100.0 & 100.0 & 99.9 & 5 & 0.059 & 0.737 & 0.546 & -0.017 & 0.485 & 0.236 \\ 
    &   & $\widehat{Q}_{\rightarrow T}$ & 100.0 & 100.0 & 97.7 & 3 & -0.035 & 0.423 & 0.180 & -0.028 & 0.355 & 0.127 \\ 
    &   & $ \widehat{X}_{\rightarrow Y}$ & 100.0 & 100.0 & 100.0 & 5 & -0.028 & 0.336 & 0.114 & -0.023 & 0.090 & 0.009 \\ 
    &   & $\widehat{Z}_{\rightarrow Y}$ & 100.0 & 100.0 & 99.8 & 3 & -0.034 & 0.368 & 0.136 & -0.027 & 0.304 & 0.093 \\ 
    &   & $\widehat{X}_{\rightarrow T, Y}$ & 100.0 & 100.0 & 99.9 & 8 & 0.045 & 0.702 & 0.495 & -0.024 & 0.119 & 0.015 \\
\hline
     \end{tabular}
 \end{minipage}
\vspace{6pt}
 \end{table} 

\clearpage

\begin{table}
\centering
 \def\~{\hphantom{0}}
 \caption{Setting 2, linear outcome model, $n=500, 1000$. The proportion (\%) of times the selected subset, $\widehat{S}$, satisfies three conditions of unconfoundedness and the median cardinality of $\widehat{S}$ (\#). Average bias (Bias), standard deviation (SD) and mean square error (MSE) from estimating ACE with PSM and TMLE.  }
 \label{Tab:2a}
 \begin{tabular}{lllrrrrrrrrrr}
\hline
\hline
\rule{0pt}{4ex}  
 & \multicolumn{6}{c}{} & \multicolumn{3}{c}{PSM} &  \multicolumn{3}{c}{TMLE}  \\ 
   \cline{8-10}  \cline{11-13}
  $n$& \multicolumn{1}{c}{Method} & \multicolumn{1}{c}{$\widehat{S}$} &  \multicolumn{1}{c}{$Y_t \perp\!\!\!\perp T \mid \widehat{S}$} & \multicolumn{1}{c}{$S  \subseteq \widehat{S}$} & \multicolumn{1}{c}{$S=\widehat{S}$} &\multicolumn{1}{c}{ \#} & \multicolumn{1}{c}{Bias} & \multicolumn{1}{c}{SD} & \multicolumn{1}{c}{MSE} & \multicolumn{1}{c}{Bias} & \multicolumn{1}{c}{SD} & \multicolumn{1}{c}{MSE} \\ 
 \hline
500 &  &$X$ & 0.0 & 100.0 & 0.0 & 100 & -2.263 & 2.141 & 9.702 & -1.819 & 0.580 & 3.643 \\ 
  \cline{2-13}
 & MMPC & $\widehat{X}_{\rightarrow T}$ & 64.6 & 63.5 & 63.5 & 5 & 0.374 & 1.604 & 2.714 & 0.334 & 1.388 & 2.037 \\ 
    &   & $\widehat{Q}_{\rightarrow T}$ & 0.2 & 0.2 & 0.2 & 2 & 2.180 & 1.324 & 6.503 & 2.180 & 1.297 & 6.436 \\ 
    &   & $ \widehat{X}_{\rightarrow Y}$ & 0.0 & 0.0 & 0.0 & 3 & 1.520 & 1.307 & 4.016 & 1.534 & 1.258 & 3.934 \\ 
    &   & $\widehat{Z}_{\rightarrow Y}$ & 0.0 & 0.6 & 0.2 & 2 & 1.884 & 1.402 & 5.515 & 1.898 & 1.362 & 5.457 \\ 
    &   & $\widehat{X}_{\rightarrow T, Y}$ & 0.0 & 1.1 & 0.0 & 7 & -2.032 & 1.415 & 6.131 & -2.133 & 0.983 & 5.517 \\
\hline   
      \end{tabular}
\vspace{6pt}
 \end{table}

 \clearpage

\begin{table}
\centering
 \def\~{\hphantom{0}}
 \caption{Setting 2, linear outcome model, $n=2000, 10000$. The proportion (\%) of times the selected subset, $\widehat{S}$, satisfies three conditions of unconfoundedness and the median cardinality of $\widehat{S}$ (\#). Average bias (Bias), standard deviation (SD) and mean square error (MSE) from estimating ACE with PSM and TMLE. }
 \label{Tab:2b}
 \begin{tabular}{lllrrrrrrrrrr}
\hline
\hline
\rule{0pt}{4ex}  
 & \multicolumn{6}{c}{} & \multicolumn{3}{c}{PSM} &  \multicolumn{3}{c}{TMLE}  \\ 
   \cline{8-10}  \cline{11-13}
  $n$& \multicolumn{1}{c}{Method} & \multicolumn{1}{c}{$\widehat{S}$} &  \multicolumn{1}{c}{$Y_t \perp\!\!\!\perp T \mid \widehat{S}$} & \multicolumn{1}{c}{$S  \subseteq \widehat{S}$} & \multicolumn{1}{c}{$S=\widehat{S}$} &\multicolumn{1}{c}{ \#} & \multicolumn{1}{c}{Bias} & \multicolumn{1}{c}{SD} & \multicolumn{1}{c}{MSE} & \multicolumn{1}{c}{Bias} & \multicolumn{1}{c}{SD} & \multicolumn{1}{c}{MSE} \\ 
  \hline
 2000& & $X$ & 0.0 & 100.0 & 0.0 & 100 & -2.706 & 0.850 & 8.044 & -2.735 & 0.352 & 7.606 \\ 
   \cline{2-13}
 & MMPC & $\widehat{X}_{\rightarrow T}$ & 15.5 & 100.0 & 15.5 & 6 & -2.349 & 1.133 & 6.802 & -2.436 & 1.027 & 6.987 \\ 
    &   & $\widehat{Q}_{\rightarrow T}$ & 7.7 & 40.1 & 7.7 & 4 & -1.304 & 1.108 & 2.929 & -1.290 & 1.071 & 2.812 \\ 
    &   & $ \widehat{X}_{\rightarrow Y}$ & 0.0 & 17.3 & 0.0 & 5 & -0.975 & 0.645 & 1.367 & -0.957 & 0.597 & 1.273 \\ 
    &   & $\widehat{Z}_{\rightarrow Y}$ & 0.1 & 83.9 & 0.3 & 4 & -0.976 & 0.684 & 1.421 & -0.956 & 0.634 & 1.315 \\ 
    &   & $\widehat{X}_{\rightarrow T, Y}$ & 0.0 & 22.0 & 0.0 & 8 & -2.743 & 0.726 & 8.051 & -2.828 & 0.369 & 8.133 \\
\hline 
     \end{tabular}
\vspace{6pt}
 \end{table}

  \clearpage

\begin{table}
\centering
 \def\~{\hphantom{0}}
 \begin{minipage}{175mm}
 \caption{Setting 2, binary outcome model, $n=500, 1000$. The proportion (\%) of times the selected subset, $\widehat{S}$, satisfies three conditions of unconfoundedness and the median cardinality of $\widehat{S}$ (\#). Average bias (Bias), standard deviation (SD) and mean square error (MSE) from estimating ACE with PSM and TMLE.  }
 \label{Tab:2e}
 \begin{tabular}{lllrrrrrrrrrr}
\hline
\hline
\rule{0pt}{4ex} 
 & \multicolumn{6}{c}{} & \multicolumn{3}{c}{PSM} &  \multicolumn{3}{c}{TMLE}  \\ 
   \cline{8-10}  \cline{11-13}
  $n$& \multicolumn{1}{c}{Method} & \multicolumn{1}{c}{$\widehat{S}$} &  \multicolumn{1}{c}{$Y_t \perp\!\!\!\perp T \mid \widehat{S}$} & \multicolumn{1}{c}{$S  \subseteq \widehat{S}$} & \multicolumn{1}{c}{$S=\widehat{S}$} &\multicolumn{1}{c}{ \#} & \multicolumn{1}{c}{Bias} & \multicolumn{1}{c}{SD} & \multicolumn{1}{c}{MSE} & \multicolumn{1}{c}{Bias} & \multicolumn{1}{c}{SD} & \multicolumn{1}{c}{MSE} \\ 
  \hline
   500&  &  $X$ & 0.0 & 100.0 & 0.0 & 100 & 0.095 & 0.103 & 0.019 & 0.015 & 0.080 & 0.007 \\
  \cline{2-13}
 & MMPC & $\widehat{X}_{\rightarrow T}$ & 64.6 & 63.5 & 63.5 & 5 & -0.014 & 0.080 & 0.007 & -0.097 & 0.173 & 0.039 \\ 
    &   & $\widehat{Q}_{\rightarrow T}$ & 0.3 & 0.3 & 0.3 & 2 & -0.086 & 0.071 & 0.012 & -0.122 & 0.173 & 0.045 \\ 
    &   & $ \widehat{X}_{\rightarrow Y}$ & 0.0 & 0.0 & 0.0 & 3 & -0.059 & 0.068 & 0.008 & -0.065 & 0.068 & 0.009 \\ 
    &   & $\widehat{Z}_{\rightarrow Y}$ & 0.0 & 1.0 & 0.7 & 2 & -0.074 & 0.071 & 0.011 & -0.081 & 0.133 & 0.024 \\ 
    &   & $\widehat{X}_{\rightarrow T, Y}$ & 0.0 & 0.3 & 0.0 & 7 & 0.083 & 0.067 & 0.011 & 0.022 & 0.113 & 0.013 \\
\hline   
     \end{tabular}
 \end{minipage}
\vspace{6pt}
 \end{table}
 
 \clearpage
 
\begin{table}
\centering
 \def\~{\hphantom{0}}
 \begin{minipage}{175mm}
 \caption{Setting 2, binary outcome model, $n=2000, 10000$. The proportion (\%) of times the selected subset, $\widehat{S}$, satisfies three conditions of unconfoundedness and the median cardinality of $\widehat{S}$ (\#). Average bias (Bias), standard deviation (SD) and mean square error (MSE) from estimating ACE with PSM and TMLE. }
 \label{Tab:2e}
 \begin{tabular}{lllrrrrrrrrrr}
\hline
\hline
\rule{0pt}{4ex}  
 & \multicolumn{6}{c}{} & \multicolumn{3}{c}{PSM} &  \multicolumn{3}{c}{TMLE}  \\ 
   \cline{8-10}  \cline{11-13}
  $n$& \multicolumn{1}{c}{Method} & \multicolumn{1}{c}{$\widehat{S}$} &  \multicolumn{1}{c}{$Y_t \perp\!\!\!\perp T \mid \widehat{S}$} & \multicolumn{1}{c}{$S  \subseteq \widehat{S}$} & \multicolumn{1}{c}{$S=\widehat{S}$} &\multicolumn{1}{c}{ \#} & \multicolumn{1}{c}{Bias} & \multicolumn{1}{c}{SD} & \multicolumn{1}{c}{MSE} & \multicolumn{1}{c}{Bias} & \multicolumn{1}{c}{SD} & \multicolumn{1}{c}{MSE} \\ 
  \hline
 2000 &  & $X$ & 0.0 & 100.0 & 0.0 & 100 & 0.105 & 0.035 & 0.012 & 0.049 & 0.086 & 0.010 \\ 
     \cline{2-13}
 & MMPC & $\widehat{X}_{\rightarrow T}$ & 15.5 & 100.0 & 15.5 & 6 & 0.090 & 0.047 & 0.010 & 0.047 & 0.140 & 0.022 \\ 
    &   & $\widehat{Q}_{\rightarrow T}$ & 6.9 & 38.4 & 6.8 & 4 & 0.049 & 0.046 & 0.005 & 0.011 & 0.110 & 0.012 \\ 
    &   & $ \widehat{X}_{\rightarrow Y}$ & 0.0 & 26.7 & 0.0 & 5 & 0.040 & 0.033 & 0.003 & 0.029 & 0.030 & 0.002 \\ 
    &   & $\widehat{Z}_{\rightarrow Y}$ & 0.1 & 77.1 & 0.6 & 4 & 0.040 & 0.035 & 0.003 & 0.027 & 0.038 & 0.002 \\ 
    &   & $\widehat{X}_{\rightarrow T, Y}$ & 0.0 & 35.3 & 0.0 & 8 & 0.104 & 0.031 & 0.012 & 0.040 & 0.117 & 0.015 \\
\hline  
    \end{tabular}
 \end{minipage}
\vspace{6pt}
 \end{table}

 \clearpage

\begin{table}
\centering
 \def\~{\hphantom{0}}
 \begin{minipage}{175mm}
 \caption{Setting 2, nonlinear outcome model, $n=500, 1000$. The proportion (\%) of times the selected subset, $\widehat{S}$, satisfies three conditions of unconfoundedness and the median cardinality of $\widehat{S}$ (\#). Average bias (Bias), standard deviation (SD) and mean square error (MSE) from estimating ACE with PSM and TMLE. }
 \label{Tab:2c}
 \begin{tabular}{lllrrrrrrrrrr}
\hline
\hline
\rule{0pt}{4ex}  
 & \multicolumn{6}{c}{} & \multicolumn{3}{c}{PSM} &  \multicolumn{3}{c}{TMLE}  \\ 
   \cline{8-10}  \cline{11-13}
  $n$& \multicolumn{1}{c}{Method} & \multicolumn{1}{c}{$\widehat{S}$} &  \multicolumn{1}{c}{$Y_t \perp\!\!\!\perp T \mid \widehat{S}$} & \multicolumn{1}{c}{$S  \subseteq \widehat{S}$} & \multicolumn{1}{c}{$S=\widehat{S}$} &\multicolumn{1}{c}{ \#} & \multicolumn{1}{c}{Bias} & \multicolumn{1}{c}{SD} & \multicolumn{1}{c}{MSE} & \multicolumn{1}{c}{Bias} & \multicolumn{1}{c}{SD} & \multicolumn{1}{c}{MSE} \\ 
  \hline    
     500&   & $X$ & 0.0 & 100.0 & 0.0 & 100 & -2.298 & 2.541 & 11.738 & -1.972 & 0.664 & 4.329 \\ 
      \cline{2-13}
  & MMPC & $\widehat{X}_{\rightarrow T}$ & 64.6 & 63.5 & 63.5 & 5 & 0.385 & 1.739 & 3.171 & 0.231 & 1.479 & 2.241 \\ 
    &   & $\widehat{Q}_{\rightarrow T}$ & 0.9 & 0.9 & 0.9 & 2 & 1.773 & 1.332 & 4.920 & 1.790 & 1.249 & 4.763 \\ 
    &   & $ \widehat{X}_{\rightarrow Y}$ & 0.0 & 0.0 & 0.0 & 3 & 0.455 & 1.389 & 2.136 & 0.480 & 1.252 & 1.798 \\ 
    &   & $\widehat{Z}_{\rightarrow Y}$ & 0.0 & 7.0 & 2.7 & 2 & 0.964 & 1.502 & 3.187 & 0.954 & 1.427 & 2.947 \\ 
    &   & $\widehat{X}_{\rightarrow T, Y}$ & 0.0 & 0.2 & 0.0 & 7 & -2.122 & 1.589 & 7.029 & -2.313 & 1.061 & 6.476 \\
\hline
        \end{tabular}
 \end{minipage}
\vspace{6pt}
 \end{table}
 
 \clearpage

\begin{table}
\centering
 \def\~{\hphantom{0}}
 \begin{minipage}{175mm}
 \caption{Setting 2, nonlinear outcome model, $n=2000, 10000$. The proportion (\%) of times the selected subset, $\widehat{S}$, satisfies three conditions of unconfoundedness and the median cardinality of $\widehat{S}$ (\#). Average bias (Bias), standard deviation (SD) and mean square error (MSE) from estimating ACE with PSM and TMLE.  }
 \label{Tab:2d}
 \begin{tabular}{lllrrrrrrrrrr}
\hline
\hline
\rule{0pt}{4ex}  
 & \multicolumn{6}{c}{} & \multicolumn{3}{c}{PSM} &  \multicolumn{3}{c}{TMLE}  \\ 
   \cline{8-10}  \cline{11-13}
  $n$& \multicolumn{1}{c}{Method} & \multicolumn{1}{c}{$\widehat{S}$} &  \multicolumn{1}{c}{$Y_t \perp\!\!\!\perp T \mid \widehat{S}$} & \multicolumn{1}{c}{$S  \subseteq \widehat{S}$} & \multicolumn{1}{c}{$S=\widehat{S}$} &\multicolumn{1}{c}{ \#} & \multicolumn{1}{c}{Bias} & \multicolumn{1}{c}{SD} & \multicolumn{1}{c}{MSE} & \multicolumn{1}{c}{Bias} & \multicolumn{1}{c}{SD} & \multicolumn{1}{c}{MSE} \\ 
  \hline    
    2000 &   & $X$ & 0.0 & 100.0 & 0.0 & 100 & -2.727 & 0.995 & 8.429 & -2.836 & 0.367 & 8.178 \\ 
       \cline{2-13}
    & MMPC & $\widehat{X}_{\rightarrow T}$ & 15.5 & 100.0 & 15.5 & 6 & -2.357 & 1.229 & 7.065 & -2.513 & 1.065 & 7.451 \\ 
    &   & $\widehat{Q}_{\rightarrow T}$ & 6.4 & 33.3 & 6.4 & 4 & -1.194 & 1.133 & 2.711 & -1.224 & 1.090 & 2.686 \\ 
    &   & $ \widehat{X}_{\rightarrow Y}$ & 0.0 & 8.9 & 0.0 & 5 & -1.146 & 0.631 & 1.711 & -1.168 & 0.522 & 1.636 \\ 
    &   & $\widehat{Z}_{\rightarrow Y}$ & 0.1 & 96.9 & 1.0 & 4 & -1.136 & 0.664 & 1.731 & -1.149 & 0.588 & 1.666 \\ 
    &   & $\widehat{X}_{\rightarrow T, Y}$ & 0.0 & 8.9 & 0.0 & 8 & -2.757 & 0.851 & 8.323 & -2.891 & 0.429 & 8.542 \\
\hline  
       \end{tabular}
 \end{minipage}
\vspace{6pt}
 \end{table} 

\clearpage

\begin{table}
\centering
 \def\~{\hphantom{0}}
 \caption{Setting 1, linear outcome model, $n=500, 1000$. The proportion (\%) of times the selected subset, $\widehat{S}$, upholds unconfoundedness and coverage probability (CP), width of confidence interval (CIW) as well as average lower and upper limits (CIL, CIU) of confidence intervals when estimating ACE by PSM and TMLE. }
 \label{Tab:3a}
 \begin{tabular}{lllrrrrrrrrr}
\hline
\hline
\rule{0pt}{4ex}  
& \multicolumn{3}{c}{} & \multicolumn{4}{c}{PSM} &  \multicolumn{4}{c}{TMLE}  \\ 
   \cline{5-12} 
 n & Method & $\widehat{S}$ &  $Y_t \perp\!\!\!\perp T \mid \widehat{S}$ & CP & CIW & CIL & CIU & CP & CIW & CIL & CIU \\
  \hline
 500 &  & $X$ & 100.0 & 97.3 & 5.7 & -0.492 & 5.172 & 49.0 & 0.3 & 1.986 & 2.279 \\ 
  \cline{2-12}
    & MMPC & $\widehat{X}_{\rightarrow T}$ & 88.1 & 93.1 & 3.5 & 0.486 & 4.016 & 65.9 & 1.4 & 1.490 & 2.925 \\ 
    &   & $\widehat{Q}_{\rightarrow T}$ & 63.6 & 75.3 & 2.1 & 1.364 & 3.448 & 65.4 & 1.5 & 1.649 & 3.159 \\ 
    &   & $ \widehat{X}_{\rightarrow Y}$ & 88.5 & 89.5 & 1.8 & 1.279 & 3.111 & 74.7 & 0.7 & 1.853 & 2.525 \\ 
    &   & $\widehat{Z}_{\rightarrow Y}$ & 84.9 & 85.8 & 1.8 & 1.278 & 3.111 & 80.4 & 1.4 & 1.513 & 2.882 \\ 
    &   & $\widehat{X}_{\rightarrow T, Y}$ & 99.2 & 98.1 & 3.6 & 0.377 & 3.975 & 66.7 & 0.6 & 1.797 & 2.406 \\
\hline  
         \end{tabular}
\vspace{6pt}
 \end{table}

 \clearpage

\begin{table}
\centering
 \def\~{\hphantom{0}}
 \caption{Setting 1, linear outcome model, $n=2000, 10000$. The proportion (\%) of times the selected subset, $\widehat{S}$, upholds unconfoundedness and coverage probability (CP), width of confidence interval (CIW) as well as average lower and upper limits (CIL, CIU) of confidence intervals when estimating ACE by PSM and TMLE. }
 \label{Tab:3b}
 \begin{tabular}{lllrrrrrrrrr}
\hline
\hline
\rule{0pt}{4ex}  
& \multicolumn{3}{c}{} & \multicolumn{4}{c}{PSM} &  \multicolumn{4}{c}{TMLE}  \\ 
   \cline{5-12} 
 n & Method & $\widehat{S}$ &  $Y_t \perp\!\!\!\perp T \mid \widehat{S}$ & CP & CIW & CIL & CIU & CP & CIW & CIL & CIU \\
 \hline
 2000 &   &  $X$ & 100.0 & 98.2 & 2.2 & 0.967 & 3.212 & 65.7 & 0.2 & 1.937 & 2.145 \\ 
   \cline{2-12}
   & MMPC & $\widehat{X}_{\rightarrow T}$ & 100.0 & 97.3 & 2.0 & 1.068 & 3.057 & 80.8 & 0.9 & 1.572 & 2.505 \\ 
    &   & $\widehat{Q}_{\rightarrow T}$ & 99.8 & 95.2 & 1.1 & 1.481 & 2.565 & 91.5 & 0.8 & 1.611 & 2.442 \\ 
    &   & $ \widehat{X}_{\rightarrow Y}$ & 100.0 & 99.9 & 0.9 & 1.564 & 2.473 & 88.5 & 0.2 & 1.918 & 2.118 \\ 
    &   & $\widehat{Z}_{\rightarrow Y}$ & 100.0 & 95.4 & 0.9 & 1.553 & 2.460 & 94.1 & 0.7 & 1.658 & 2.381 \\ 
    &   & $\widehat{X}_{\rightarrow T, Y}$ & 100.0 & 98.2 & 2.0 & 1.078 & 3.080 & 76.7 & 0.2 & 1.908 & 2.145 \\
\hline
       \end{tabular}
\vspace{6pt}
 \end{table}

  \clearpage
 
\begin{table}
\centering
 \def\~{\hphantom{0}}
 \begin{minipage}{175mm}
 \caption{Setting 1, binary outcome model, $n=500, 1000$. The proportion (\%) of times the selected subset, $\widehat{S}$, upholds unconfoundedness and coverage probability (CP), width of confidence interval (CIW) as well as average lower and upper limits (CIL, CIU) of confidence intervals when estimating ACE by PSM and TMLE.  }
 \label{Tab:3c}
 \begin{tabular}{lllrrrrrrrrr}
\hline
\hline
\rule{0pt}{4ex}  
& \multicolumn{3}{c}{} & \multicolumn{4}{c}{PSM} &  \multicolumn{4}{c}{TMLE}  \\ 
   \cline{5-12} 
 n & Method & $\widehat{S}$ &  $Y_t \perp\!\!\!\perp T \mid \widehat{S}$ & CP & CIW & CIL & CIU & CP & CIW & CIL & CIU \\
  \hline
  500 &  &$X$ & 100.0 & 97.2 & 0.5 & -0.157 & 0.373 & 37.1 & 0.1 & -0.001 & 0.115 \\ 
  \cline{2-12}
    & MMPC & $\widehat{X}_{\rightarrow T}$ & 88.1 & 96.8 & 0.3 & -0.056 & 0.281 & 64.7 & 0.2 & -0.031 & 0.204 \\ 
    &   & $\widehat{Q}_{\rightarrow T}$ & 57.3 & 87.3 & 0.2 & -0.006 & 0.203 & 76.5 & 0.2 & -0.053 & 0.151 \\ 
    &   & $ \widehat{X}_{\rightarrow Y}$ & 68.4 & 86.8 & 0.2 & 0.007 & 0.193 & 73.9 & 0.1 & 0.028 & 0.151 \\ 
    &   & $\widehat{Z}_{\rightarrow Y}$ & 66.4 & 86.3 & 0.2 & 0.009 & 0.197 & 80.9 & 0.2 & -0.013 & 0.165 \\ 
    &   & $\widehat{X}_{\rightarrow T, Y}$ & 97.7 & 98.2 & 0.3 & -0.056 & 0.286 & 46.7 & 0.2 & -0.011 & 0.145 \\
\hline    
     \end{tabular}
 \end{minipage}
\vspace{6pt}
 \end{table}
 
 \clearpage
 
\begin{table}
\centering
 \def\~{\hphantom{0}}
 \begin{minipage}{175mm}
 \caption{Setting 1, binary outcome model, $n=2000, 10000$. The proportion (\%) of times the selected subset, $\widehat{S}$, upholds unconfoundedness and coverage probability (CP), width of confidence interval (CIW) as well as average lower and upper limits (CIL, CIU) of confidence intervals when estimating ACE by PSM and TMLE. }
 \label{Tab:3d}
 \begin{tabular}{lllrrrrrrrrr}
\hline
\hline
\rule{0pt}{4ex}  
& \multicolumn{3}{c}{} & \multicolumn{4}{c}{PSM} &  \multicolumn{4}{c}{TMLE}  \\ 
   \cline{5-12} 
 n & Method & $\widehat{S}$ &  $Y_t \perp\!\!\!\perp T \mid \widehat{S}$ & CP & CIW & CIL & CIU & CP & CIW & CIL & CIU \\
  \hline
  2000& & $X$ & 100.0 & 99.1 & 0.2 & 0.014 & 0.226 & 38.5 & 0.1 & -0.010 & 0.081 \\ 
      \cline{2-12}
   & MMPC & $\widehat{X}_{\rightarrow T}$ & 100.0 & 97.7 & 0.2 & 0.026 & 0.215 & 78.0 & 0.2 & 0.039 & 0.191 \\ 
    &   & $\widehat{Q}_{\rightarrow T}$ & 99.6 & 96.9 & 0.1 & 0.066 & 0.177 & 91.4 & 0.1 & 0.038 & 0.153 \\ 
    &   & $ \widehat{X}_{\rightarrow Y}$ & 100.0 & 98.9 & 0.1 & 0.074 & 0.171 & 89.2 & 0.1 & 0.091 & 0.146 \\ 
    &   & $\widehat{Z}_{\rightarrow Y}$ & 100.0 & 96.4 & 0.1 & 0.075 & 0.172 & 94.2 & 0.1 & 0.063 & 0.156 \\ 
    &   & $\widehat{X}_{\rightarrow T, Y}$ & 100.0 & 99.0 & 0.2 & 0.024 & 0.215 & 44.3 & 0.1 & -0.030 & 0.078 \\
\hline       
     \end{tabular}
 \end{minipage}
\vspace{6pt}
 \end{table}
 
 \clearpage
  
\begin{table}
\centering
 \def\~{\hphantom{0}}
 \begin{minipage}{175mm}
 \caption{Setting 1, nonlinear outcome model, $n=500, 1000$. The proportion (\%) of times the selected subset, $\widehat{S}$, upholds unconfoundedness and coverage probability (CP), width of confidence interval (CIW) as well as average lower and upper limits (CIL, CIU) of confidence intervals when estimating ACE by PSM and TMLE. }
 \label{Tab:3e}
 \begin{tabular}{lllrrrrrrrrr}
\hline
\hline
\rule{0pt}{4ex}  
& \multicolumn{3}{c}{} & \multicolumn{4}{c}{PSM} &  \multicolumn{4}{c}{TMLE}  \\ 
   \cline{5-12} 
 n & Method & $\widehat{S}$ &  $Y_t \perp\!\!\!\perp T \mid \widehat{S}$ & CP & CIW & CIL & CIU & CP & CIW & CIL & CIU \\
   \hline    
500&   &$X$ & 100.0 & 94.6 & 7.3 & -1.376 & 5.949 & 72.9 & 0.6 & 1.715 & 2.311 \\ 
     \cline{2-12}
    & MMPC & $\widehat{X}_{\rightarrow T}$ & 88.1 & 92.1 & 4.6 & -0.033 & 4.584 & 61.7 & 1.8 & 1.171 & 3.000 \\ 
    &   & $\widehat{Q}_{\rightarrow T}$ & 83.2 & 85.2 & 2.8 & 0.751 & 3.594 & 76.9 & 2.0 & 1.133 & 3.156 \\ 
    &   & $ \widehat{X}_{\rightarrow Y}$ & 99.0 & 92.9 & 2.6 & 0.683 & 3.251 & 88.3 & 1.1 & 1.408 & 2.555 \\ 
    &   & $\widehat{Z}_{\rightarrow Y}$ & 95.2 & 91.8 & 2.5 & 0.748 & 3.297 & 85.2 & 1.9 & 1.071 & 2.949 \\ 
    &   & $\widehat{X}_{\rightarrow T, Y}$ & 99.9 & 96.8 & 4.7 & -0.194 & 4.502 & 72.8 & 1.1 & 1.412 & 2.486 \\
\hline 
    \end{tabular}
 \end{minipage}
\vspace{6pt}
 \end{table}
 
 \clearpage

\begin{table}
\centering
 \def\~{\hphantom{0}}
 \begin{minipage}{175mm}
 \caption{Setting 1, nonlinear outcome model, $n=2000, 10000$. The proportion (\%) of times the selected subset, $\widehat{S}$, upholds unconfoundedness and coverage probability (CP), width of confidence interval (CIW) as well as average lower and upper limits (CIL, CIU) of confidence intervals when estimating ACE by PSM and TMLE. }
 \label{Tab:3f}
 \begin{tabular}{lllrrrrrrrrr}
\hline
\hline
\rule{0pt}{4ex}  
& \multicolumn{3}{c}{} & \multicolumn{4}{c}{PSM} &  \multicolumn{4}{c}{TMLE}  \\ 
   \cline{5-12} 
 n & Method & $\widehat{S}$ &  $Y_t \perp\!\!\!\perp T \mid \widehat{S}$ & CP & CIW & CIL & CIU & CP & CIW & CIL & CIU \\
  \hline    
 2000 &   &  $X$ & 100.0 & 94.7 & 2.9 & 0.592 & 3.542 & 78.2 & 0.3 & 1.794 & 2.136 \\ 
  \cline{2-12}
     & MMPC & $\widehat{X}_{\rightarrow T}$ & 100.0 & 92.7 & 2.6 & 0.749 & 3.369 & 77.3 & 1.2 & 1.375 & 2.591 \\ 
    &   & $\widehat{Q}_{\rightarrow T}$ & 100.0 & 92.1 & 1.5 & 1.238 & 2.693 & 89.1 & 1.1 & 1.401 & 2.544 \\ 
    &   & $ \widehat{X}_{\rightarrow Y}$ & 100.0 & 93.8 & 1.3 & 1.343 & 2.601 & 91.9 & 0.3 & 1.818 & 2.137 \\ 
    &   & $\widehat{Z}_{\rightarrow Y}$ & 100.0 & 90.1 & 1.3 & 1.338 & 2.595 & 90.6 & 1.0 & 1.458 & 2.487 \\ 
    &   & $\widehat{X}_{\rightarrow T, Y}$ & 100.0 & 94.8 & 2.6 & 0.729 & 3.361 & 84.4 & 0.3 & 1.802 & 2.149 \\
\hline
     \end{tabular}
 \end{minipage}
\vspace{6pt}
 \end{table} 

\clearpage

\begin{table}
\centering
 \def\~{\hphantom{0}}
 \caption{Setting 2, linear outcome model, $n=500, 1000$. The proportion (\%) of times the selected subset, $\widehat{S}$, upholds unconfoundedness and coverage probability (CP), width of confidence interval (CIW) as well as average lower and upper limits (CIL, CIU) of confidence intervals when estimating ACE by PSM and TMLE. }
 \label{Tab:4a}
 \begin{tabular}{lllrrrrrrrrr}
\hline
\hline
\rule{0pt}{4ex}  
& \multicolumn{3}{c}{} & \multicolumn{4}{c}{PSM} &  \multicolumn{4}{c}{TMLE}  \\ 
   \cline{5-12} 
 n & Method & $\widehat{S}$ &  $Y_t \perp\!\!\!\perp T \mid \widehat{S}$ & CP & CIW & CIL & CIU & CP & CIW & CIL & CIU \\
  \hline
500 &  & $X$ & 0.0 & 83.9 & 9.0 & -4.764 & 4.239 & 1.7 & 1.1 & -0.379 & 0.742 \\ 
  \cline{2-12}
   & MMPC & $\widehat{X}_{\rightarrow T}$ & 64.6 & 91.5 & 5.7 & -0.487 & 5.236 & 71.2 & 3.0 & 0.820 & 3.848 \\ 
    &   & $\widehat{Q}_{\rightarrow T}$ & 0.2 & 40.9 & 3.6 & 2.376 & 5.984 & 30.8 & 3.0 & 2.660 & 5.700 \\ 
    &   & $ \widehat{X}_{\rightarrow Y}$ & 0.0 & 53.3 & 3.1 & 1.987 & 5.052 & 37.5 & 2.0 & 2.540 & 4.528 \\ 
    &   & $\widehat{Z}_{\rightarrow Y}$ & 0.0 & 42.1 & 3.0 & 2.373 & 5.395 & 34.5 & 2.4 & 2.679 & 5.116 \\ 
    &   & $\widehat{X}_{\rightarrow T, Y}$ & 0.0 & 68.9 & 5.7 & -2.861 & 2.798 & 12.5 & 1.7 & -0.977 & 0.710 \\
\hline   
      \end{tabular}
\vspace{6pt}
 \end{table}

 \clearpage

\begin{table}
\centering
 \def\~{\hphantom{0}}
 \caption{Setting 2, linear outcome model, $n=2000, 10000$. The proportion (\%) of times the selected subset, $\widehat{S}$, upholds unconfoundedness and coverage probability (CP), width of confidence interval (CIW) as well as average lower and upper limits (CIL, CIU) of confidence intervals when estimating ACE by PSM and TMLE.  }
 \label{Tab:4b}
 \begin{tabular}{lllrrrrrrrrr}
\hline
\hline
\rule{0pt}{4ex}  
& \multicolumn{3}{c}{} & \multicolumn{4}{c}{PSM} &  \multicolumn{4}{c}{TMLE}  \\ 
   \cline{5-12} 
 n & Method & $\widehat{S}$ &  $Y_t \perp\!\!\!\perp T \mid \widehat{S}$ & CP & CIW & CIL & CIU & CP & CIW & CIL & CIU \\
  \hline
 2000& &   $X$ & 0.0 & 15.6 & 3.6 & -2.494 & 1.081 & 0.0 & 0.8 & -1.138 & -0.333 \\ 
   \cline{2-12}
  & MMPC & $\widehat{X}_{\rightarrow T}$ & 15.5 & 21.9 & 3.2 & -1.949 & 1.252 & 12.6 & 1.4 & -1.118 & 0.247 \\ 
    &   & $\widehat{Q}_{\rightarrow T}$ & 7.7 & 36.5 & 2.2 & -0.389 & 1.781 & 19.9 & 1.3 & 0.080 & 1.340 \\ 
    &   & $ \widehat{X}_{\rightarrow Y}$ & 0.0 & 43.5 & 1.8 & 0.106 & 1.944 & 14.8 & 1.0 & 0.565 & 1.521 \\ 
    &   & $\widehat{Z}_{\rightarrow Y}$ & 0.1 & 42.7 & 1.8 & 0.106 & 1.941 & 19.9 & 1.1 & 0.479 & 1.610 \\ 
    &   & $\widehat{X}_{\rightarrow T, Y}$ & 0.0 & 8.7 & 3.2 & -2.341 & 0.855 & 0.0 & 1.0 & -1.326 & -0.330 \\
\hline 
     \end{tabular}
\vspace{6pt}
 \end{table}

  \clearpage

\begin{table}
\centering
 \def\~{\hphantom{0}}
 \begin{minipage}{175mm}
 \caption{Setting 2, binary outcome model, $n=500, 1000$. The proportion (\%) of times the selected subset, $\widehat{S}$, upholds unconfoundedness and coverage probability (CP), width of confidence interval (CIW) as well as average lower and upper limits (CIL, CIU) of confidence intervals when estimating ACE by PSM and TMLE.  }
 \label{Tab:4c}
 \begin{tabular}{lllrrrrrrrrr}
\hline
\hline
\rule{0pt}{4ex}  
& \multicolumn{3}{c}{} & \multicolumn{4}{c}{PSM} &  \multicolumn{4}{c}{TMLE}  \\ 
   \cline{5-12} 
 n & Method & $\widehat{S}$ &  $Y_t \perp\!\!\!\perp T \mid \widehat{S}$ & CP & CIW & CIL & CIU & CP & CIW & CIL & CIU \\
  \hline
   500&  &$X$ & 0.0 & 94.8 & 0.5 & -0.070 & 0.424 & 58.1 & 0.1 & 0.037 & 0.158 \\ 
  \cline{2-12}
 & MMPC & $\widehat{X}_{\rightarrow T}$ & 64.6 & 94.6 & 0.3 & -0.092 & 0.229 & 68.7 & 0.3 & -0.149 & 0.120 \\ 
    &   & $\widehat{Q}_{\rightarrow T}$ & 0.3 & 60.5 & 0.2 & -0.104 & 0.098 & 48.8 & 0.2 & -0.144 & 0.066 \\ 
    &   & $ \widehat{X}_{\rightarrow Y}$ & 0.0 & 64.9 & 0.2 & -0.070 & 0.117 & 54.1 & 0.2 & -0.057 & 0.093 \\ 
    &   & $\widehat{Z}_{\rightarrow Y}$ & 0.0 & 58.6 & 0.2 & -0.082 & 0.098 & 52.8 & 0.2 & -0.085 & 0.088 \\ 
    &   & $\widehat{X}_{\rightarrow T, Y}$ & 0.0 & 87.1 & 0.3 & 0.006 & 0.326 & 65.0 & 0.2 & 0.024 & 0.185 \\
\hline   
     \end{tabular}
 \end{minipage}
\vspace{6pt}
 \end{table}
 
 \clearpage
 
\begin{table}
\centering
 \def\~{\hphantom{0}}
 \begin{minipage}{175mm}
 \caption{Setting 2, binary outcome model, $n=2000, 10000$. The proportion (\%) of times the selected subset, $\widehat{S}$, upholds unconfoundedness and coverage probability (CP), width of confidence interval (CIW) as well as average lower and upper limits (CIL, CIU) of confidence intervals when estimating ACE by PSM and TMLE. }
 \label{Tab:4d}
  \begin{tabular}{lllrrrrrrrrr}
\hline
\hline
\rule{0pt}{4ex}  
& \multicolumn{3}{c}{} & \multicolumn{4}{c}{PSM} &  \multicolumn{4}{c}{TMLE}  \\ 
   \cline{5-12} 
 n & Method & $\widehat{S}$ &  $Y_t \perp\!\!\!\perp T \mid \widehat{S}$ & CP & CIW & CIL & CIU & CP & CIW & CIL & CIU \\
  \hline
 2000 &  &$X$ & 0.0 & 36.6 & 0.2 & 0.088 & 0.286 & 28.1 & 0.1 & 0.091 & 0.173 \\ 
     \cline{2-12}
   & MMPC & $\widehat{X}_{\rightarrow T}$ & 15.5 & 38.7 & 0.2 & 0.083 & 0.262 & 39.6 & 0.1 & 0.071 & 0.189 \\ 
    &   & $\widehat{Q}_{\rightarrow T}$ & 6.9 & 56.4 & 0.1 & 0.070 & 0.194 & 58.6 & 0.1 & 0.043 & 0.146 \\ 
    &   & $ \widehat{X}_{\rightarrow Y}$ & 0.0 & 67.1 & 0.1 & 0.069 & 0.177 & 61.5 & 0.1 & 0.073 & 0.151 \\ 
    &   & $\widehat{Z}_{\rightarrow Y}$ & 0.1 & 67.1 & 0.1 & 0.068 & 0.176 & 68.2 & 0.1 & 0.066 & 0.154 \\ 
    &   & $\widehat{X}_{\rightarrow T, Y}$ & 0.0 & 27.0 & 0.2 & 0.098 & 0.276 & 20.9 & 0.1 & 0.073 & 0.171 \\
\hline  
    \end{tabular}
 \end{minipage}
\vspace{6pt}
 \end{table}

 \clearpage

\begin{table}
\centering
 \def\~{\hphantom{0}}
 \begin{minipage}{175mm}
 \caption{Setting 2, nonlinear outcome model, $n=500, 1000$. The proportion (\%) of times the selected subset, $\widehat{S}$, upholds unconfoundedness and coverage probability (CP), width of confidence interval (CIW) as well as average lower and upper limits (CIL, CIU) of confidence intervals when estimating ACE by PSM and TMLE. }
 \label{Tab:4e}
 \begin{tabular}{lllrrrrrrrrr}
\hline
\hline
\rule{0pt}{4ex}  
& \multicolumn{3}{c}{} & \multicolumn{4}{c}{PSM} &  \multicolumn{4}{c}{TMLE}  \\ 
   \cline{5-12} 
 n & Method & $\widehat{S}$ &  $Y_t \perp\!\!\!\perp T \mid \widehat{S}$ & CP & CIW & CIL & CIU & CP & CIW & CIL & CIU \\
 \hline    
     500&   &$X$ & 0.0 & 83.3 & 9.9 & -5.260 & 4.664 & 2.3 & 1.3 & -0.614 & 0.671 \\ 
           \cline{2-12}
    & MMPC & $\widehat{X}_{\rightarrow T}$ & 64.6 & 91.7 & 6.3 & -0.740 & 5.510 & 72.7 & 3.3 & 0.589 & 3.873 \\ 
    &   & $\widehat{Q}_{\rightarrow T}$ & 0.9 & 56.4 & 4.1 & 1.717 & 5.830 & 44.7 & 3.3 & 2.130 & 5.450 \\ 
    &   & $ \widehat{X}_{\rightarrow Y}$ & 0.0 & 81.2 & 3.7 & 0.616 & 4.295 & 67.8 & 2.3 & 1.350 & 3.610 \\ 
    &   & $\widehat{Z}_{\rightarrow Y}$ & 0.0 & 69.9 & 3.7 & 1.096 & 4.833 & 60.7 & 2.8 & 1.571 & 4.336 \\ 
    &   & $\widehat{X}_{\rightarrow T, Y}$ & 0.0 & 72.4 & 6.3 & -3.252 & 3.008 & 13.2 & 2.0 & -1.312 & 0.686 \\
\hline
        \end{tabular}
 \end{minipage}
\vspace{6pt}
 \end{table}
 
 \clearpage

\begin{table}
\centering
 \def\~{\hphantom{0}}
 \begin{minipage}{175mm}
 \caption{Setting 2, nonlinear outcome model, $n=2000, 10000$. The proportion (\%) of times the selected subset, $\widehat{S}$, upholds unconfoundedness and coverage probability (CP), width of confidence interval (CIW) as well as average lower and upper limits (CIL, CIU) of confidence intervals when estimating ACE by PSM and TMLE.  }
 \label{Tab:4f}
 \begin{tabular}{lllrrrrrrrrr}
\hline
\hline
\rule{0pt}{4ex}  
& \multicolumn{3}{c}{} & \multicolumn{4}{c}{PSM} &  \multicolumn{4}{c}{TMLE}  \\ 
   \cline{5-12} 
 n & Method & $\widehat{S}$ &  $Y_t \perp\!\!\!\perp T \mid \widehat{S}$ & CP & CIW & CIL & CIU & CP & CIW & CIL & CIU \\
  \hline    
    2000 &   &$X$ & 0.0 & 21.7 & 3.9 & -2.695 & 1.241 & 0.0 & 0.9 & -1.270 & -0.403 \\ 
       \cline{2-12}
     & MMPC & $\widehat{X}_{\rightarrow T}$ & 15.5 & 27.5 & 3.5 & -2.117 & 1.404 & 11.9 & 1.6 & -1.291 & 0.265 \\ 
    &   & $\widehat{Q}_{\rightarrow T}$ & 6.4 & 46.4 & 2.4 & -0.371 & 1.982 & 25.5 & 1.4 & 0.058 & 1.493 \\ 
    &   & $ \widehat{X}_{\rightarrow Y}$ & 0.0 & 42.8 & 2.1 & -0.175 & 1.883 & 6.3 & 1.1 & 0.294 & 1.370 \\ 
    &   & $\widehat{Z}_{\rightarrow Y}$ & 0.1 & 43.6 & 2.1 & -0.166 & 1.894 & 16.2 & 1.3 & 0.191 & 1.512 \\ 
    &   & $\widehat{X}_{\rightarrow T, Y}$ & 0.0 & 13.9 & 3.5 & -2.516 & 1.003 & 0.0 & 1.1 & -1.459 & -0.323 \\
\hline  
       \end{tabular}
 \end{minipage}
\vspace{6pt}
 \end{table} 

\clearpage


\section*{Appendix C: Table and Figures from Section 6}
\label{App:4}

\begin{table}
 \caption{Distribution of demographic and perinatal factors by mode of delivery. The first antenatal visit takes place around pregnancy week 12. Median disposable income in the total Swedish population was 184700SEK in the year 2005.}
\begin{center}
\begin{tabular}{lrrr}
\hline
\hline
Covariate & \multicolumn{1}{c}{Total} & \multicolumn{1}{c}{C-section} & \multicolumn{1}{c}{Vaginal} \\ 
& \multicolumn{1}{c}{($n=20867$), \%}&  \multicolumn{1}{c}{($n=4356$), \%}&\multicolumn{1}{c}{($n=16511$), \%}\\ 
 \hline
  
Maternal age at delivery & & & \\
\hspace{0.4cm}$\leq 25$&23.5 &14.5 &25.9 \\
\hspace{0.4cm}26 to 35& 67.8& 70.2& 67.2\\
\hspace{0.4cm}$\geq 36$& 8.7& 15.3& 6.9\\
Maternal asthma & & & \\
\hspace{0.4cm}Yes& 9.4& 10.0&9.1 \\
\hline 
\end{tabular}
\end{center}
\end{table}
\clearpage
\setcounter{table}{24}

\begin{table}
 \caption{Continued.}
\begin{center}
\begin{tabular}{lrrr}
\hline
\hline
Covariate & \multicolumn{1}{c}{Total} & \multicolumn{1}{c}{C-section} & \multicolumn{1}{c}{Vaginal} \\ 
& \multicolumn{1}{c}{($n=20867$), \%}&  \multicolumn{1}{c}{($n=4356$), \%}&\multicolumn{1}{c}{($n=16511$), \%}\\ 
 \hline
 Paternal highest education year 2005& & & \\
\hspace{0.4cm}$\leq$ 9 years&9.7 &8.6 & 9.9\\
\hspace{0.4cm}10 to 12 years&58.9 &58.8 &59.0 \\
\hspace{0.4cm}$\geq$ 13 years&31.4& 32.6& 31.1\\
Paternal disposable income year 2005 & & & \\
\hspace{0.4cm} No income&0.7&0.7 & 0.7\\
\hspace{0.4cm}$\leq$ 184700 SEK &44.4 &41.1 & 45.3\\
\hspace{0.4cm}$>$ 184700SEK& 54.8& 58.2& 54.0\\ 
Paternal unemployment benefit year 2005& & & \\
\hspace{0.4cm}Received&12.1 & 10.8&12.4 \\
\hspace{0.4cm}Not received& 87.9&89.2&87.6\\ 
Paternal sickness benefit year 2005& & & \\
\hspace{0.4cm}Received& 6.4& 6.4&6.3 \\
\hspace{0.4cm}Not received& 93.6&93.6 &93.7 \\
Offspring sex& & & \\
\hspace{0.4cm}Female&50.0 &47.9 & 50.6\\
\hspace{0.4cm}Male&50.0& 52.1&49.4 \\
Gestational age (GA) & & & \\
\hspace{0.4cm}37 weeks & 5.2&6.1 & 5.0\\
\hspace{0.4cm}38 to 40 weeks&65.5&59.0 & 67.2\\
\hspace{0.4cm}$\geq 41$ weeks& 29.2& 34.9& 27.7\\
Birth weight & & & \\
\hspace{0.4cm}$< 2500$g& 1.1&2.2 &0.9 \\
\hspace{0.4cm}2500g to 4500g&96.1& 91.3&97.3 \\
\hspace{0.4cm}$> 4500$g& 2.8& 6.5& 1.8\\
Small for gestational age (SGA)& & & \\
\hspace{0.4cm}weight $<$ 10th percentile for the gestational age& 2.2& 3.3& 1.9\\
\hspace{0.4cm}Other& 97.8& 96.7&98.1 \\
Large for gestational age (LGA)& & & \\
\hspace{0.4cm}weight $>$ 90th percentile for the gestational age&2.0&4.9 & 1.2\\
\hspace{0.4cm}Other& 98.0&95.1 &98.8 \\
Birth place& & & \\
\hspace{0.4cm}Northern northern Sweden& 6.1& 5.1& 6.3\\
\hspace{0.4cm}Southern northern Sweden& 7.6&6.9 &7.8 \\
\hspace{0.4cm}South east Sweden&38.4& 40.5& 37.8\\
\hspace{0.4cm}South west Sweden& 48.0& 47.5&48.1 \\
\hline 
\end{tabular}
\end{center}
\end{table}

\clearpage

\begin{figure}[h]
\centerline{
\includegraphics[width=0.6\textwidth]{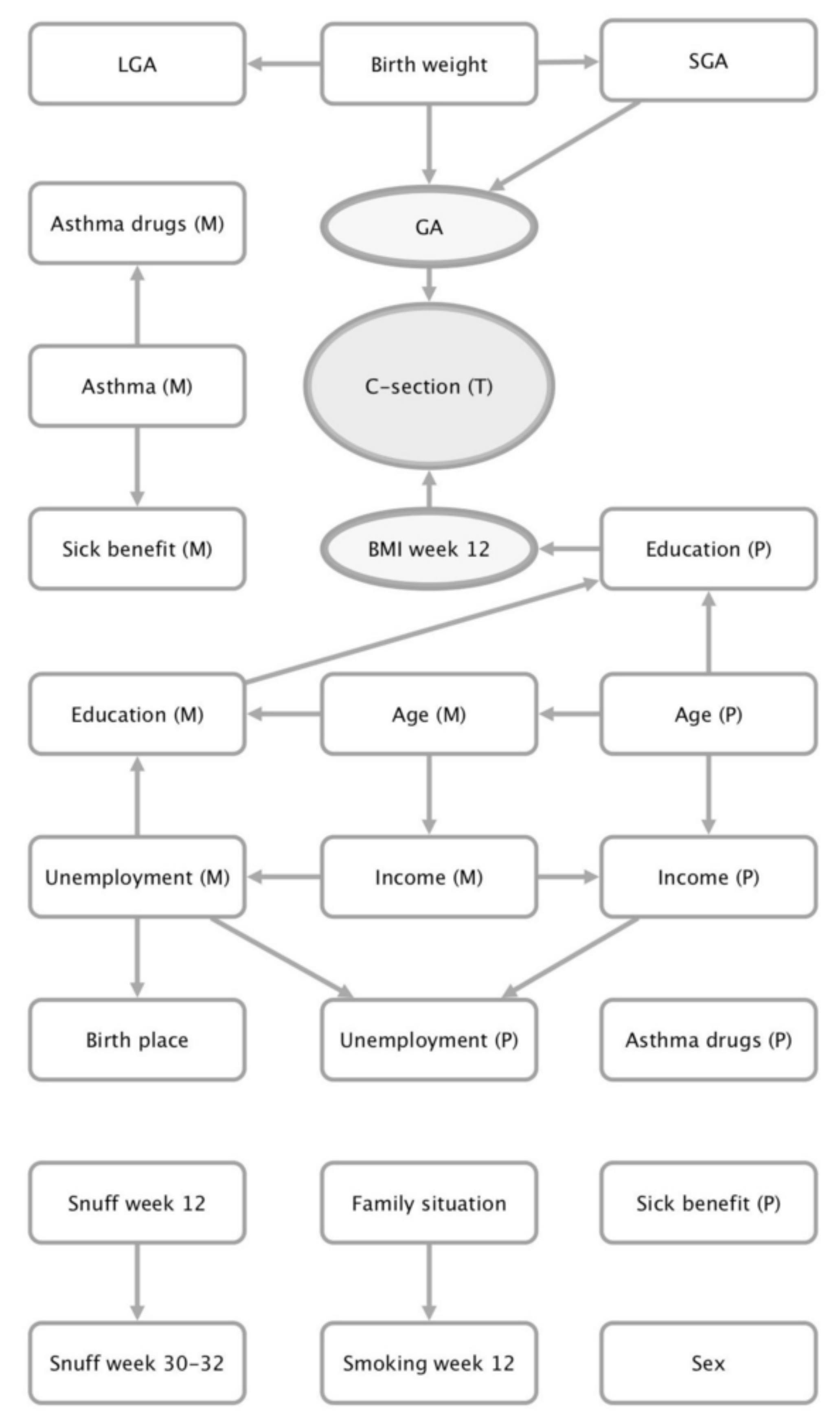}}
\captionsetup{textfont=normal, labelsep=period}
\caption{DAG resulting from MMHC, $\widehat{X}_{\rightarrow T}$.}
\label{Fig:4}
\end{figure}

\clearpage

\begin{figure}[h]
\centerline{
\includegraphics[width=0.6\textwidth]{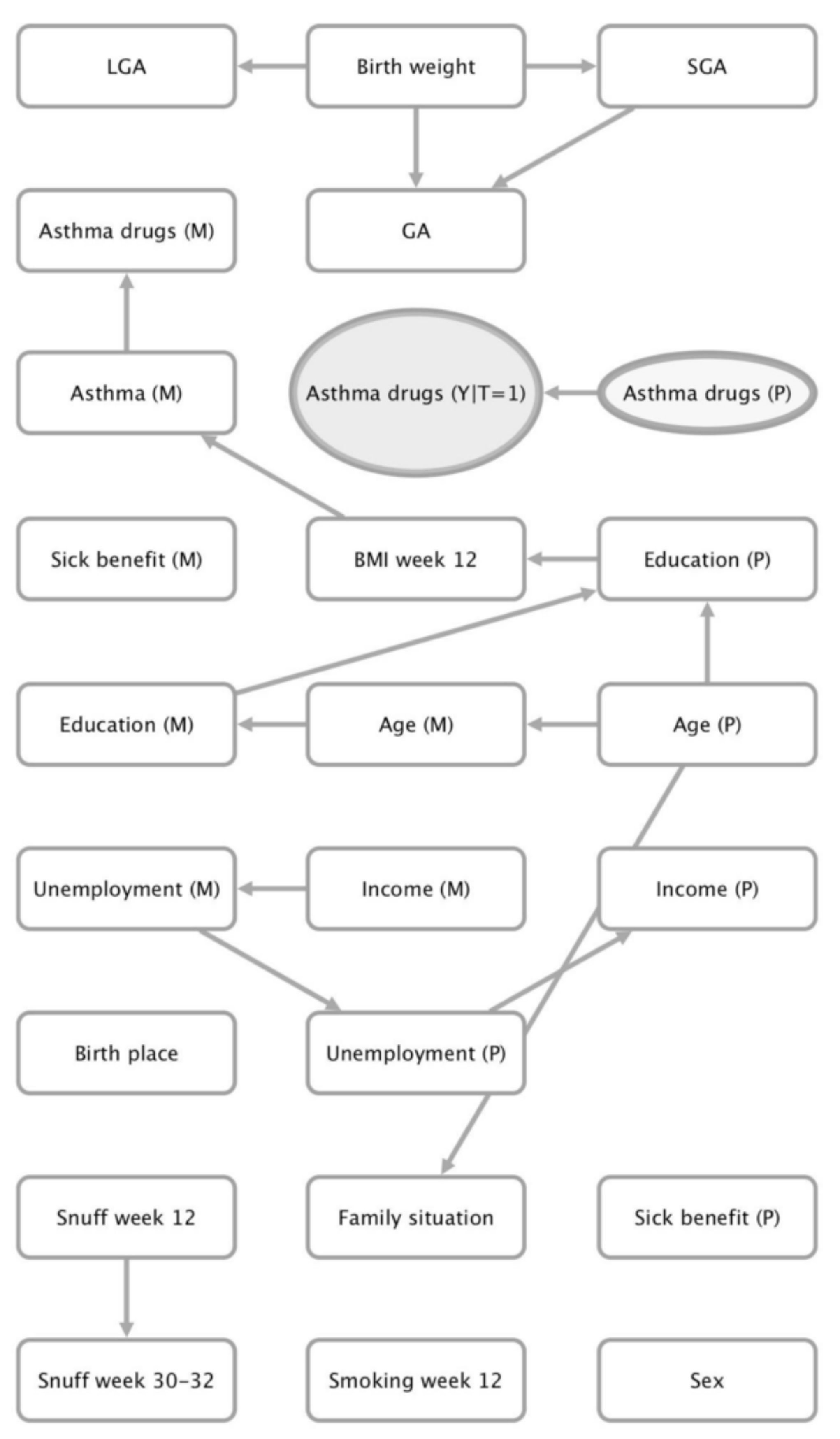}}
\captionsetup{textfont=normal, labelsep=period}
\caption{DAG resulting from MMHC, $ \widehat{X}_{\rightarrow Y}^1$}
\label{Fig:7}
\end{figure}

\clearpage

\begin{figure}[h]
\centerline{
\includegraphics[width=0.6\textwidth]{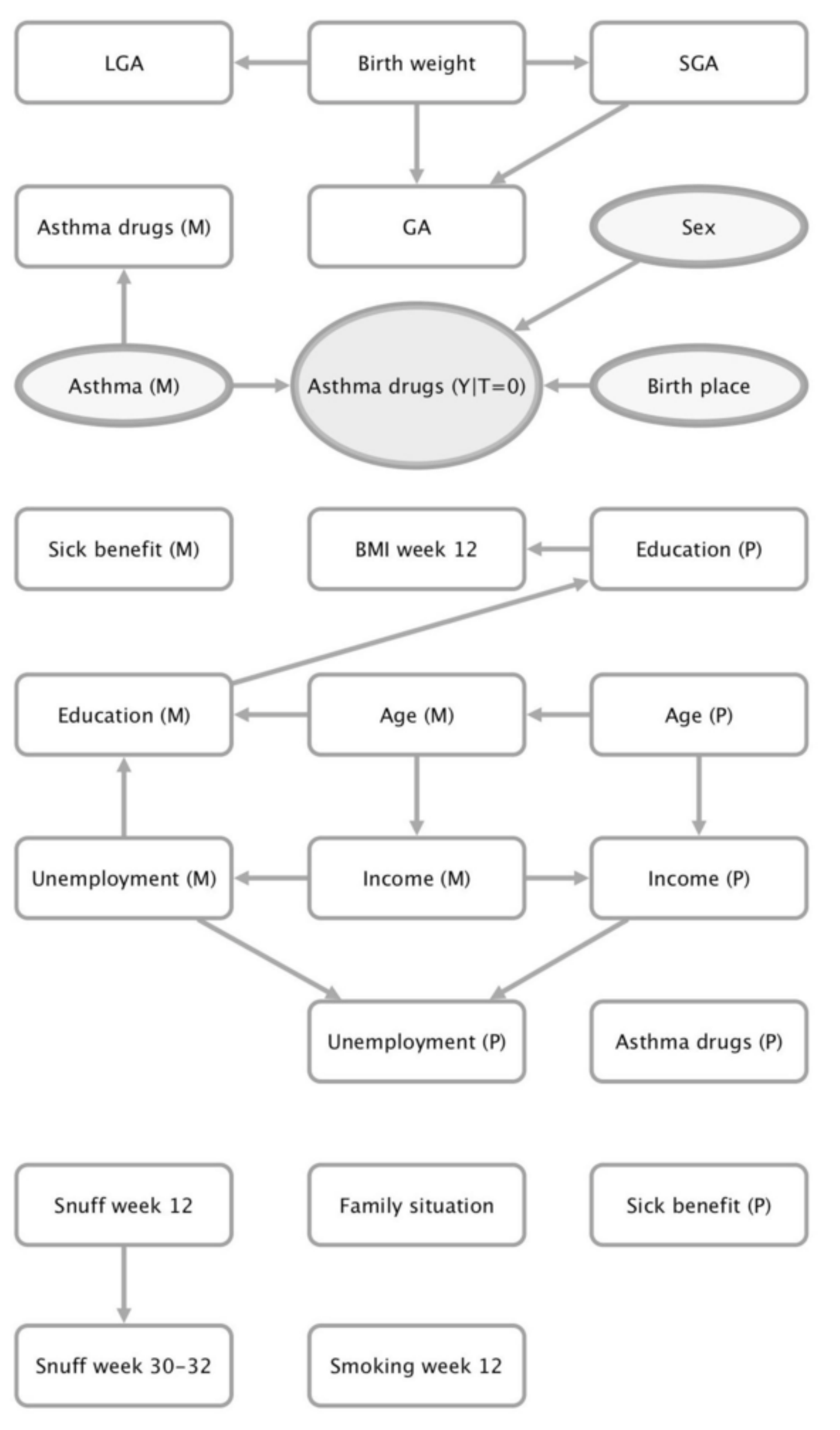}}
\captionsetup{textfont=normal, labelsep=period}
\caption{DAG resulting from MMHC,  $ \widehat{X}_{\rightarrow Y}^0$}
\label{Fig:8}
\end{figure}

\clearpage
\begin{figure}[h]
\begin{minipage}{0.5\textwidth}
\centerline{
\includegraphics[width=0.76\textwidth]{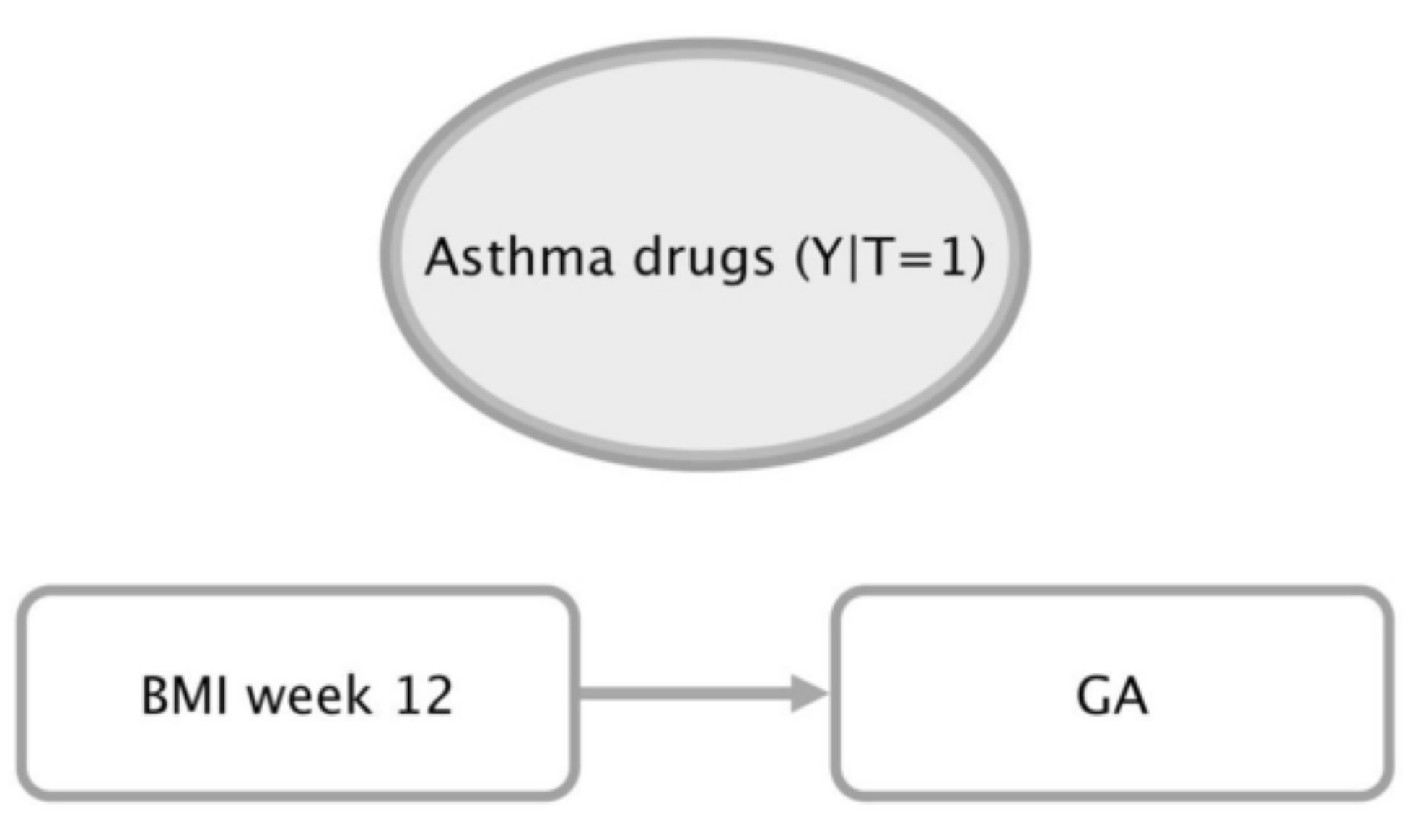}}
\captionsetup{textfont=normal, labelsep=period}
\caption{DAG resulting from MMHC,  $ \widehat{Q}_{\rightarrow T}^1$}
\label{Fig:5}
\end{minipage}
\begin{minipage}{0.5\textwidth}
\centerline{
\includegraphics[width=0.76\textwidth]{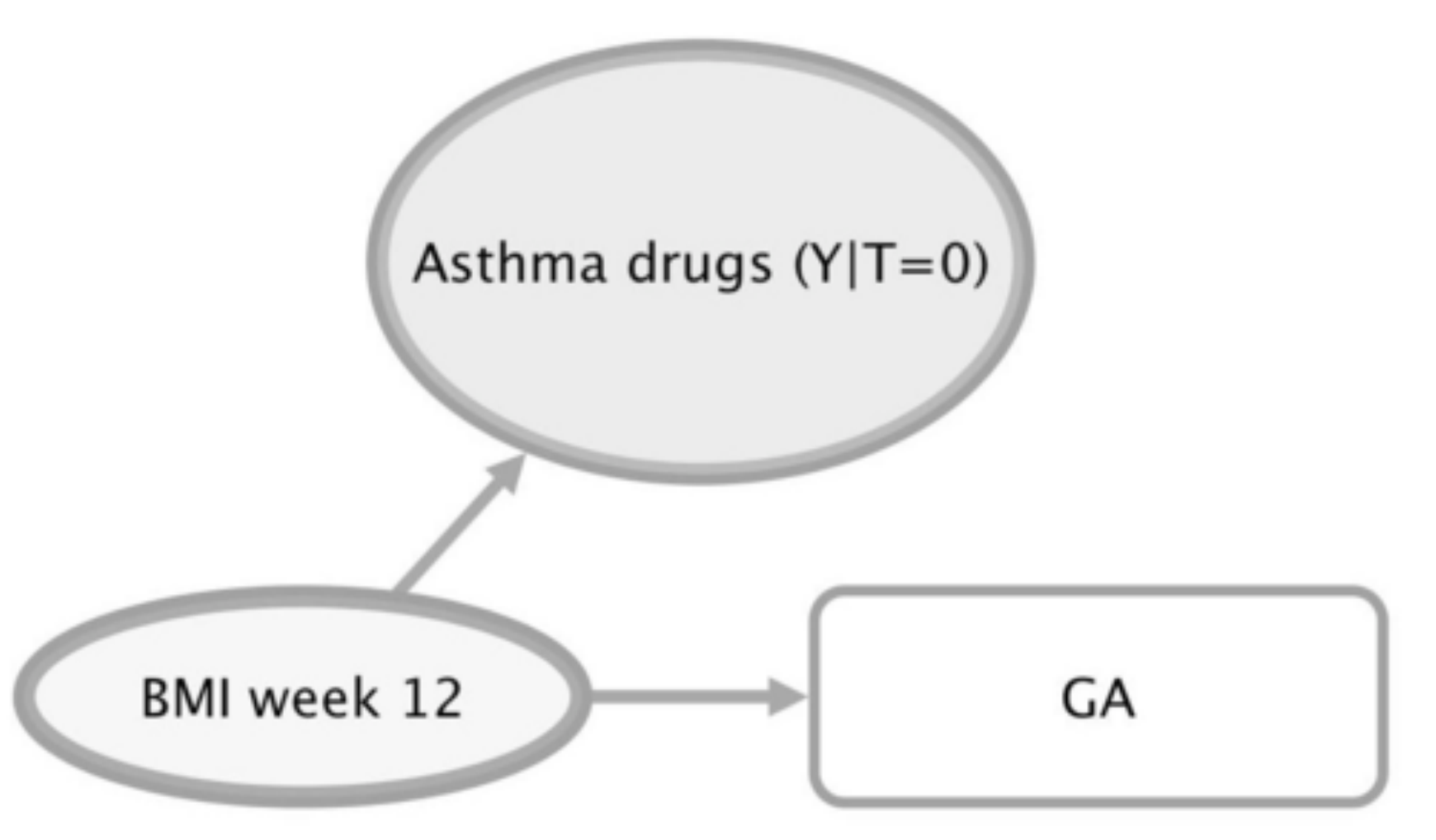}}
\captionsetup{textfont=normal, labelsep=period}
\caption{DAG resulting from MMHC, $ \widehat{Q}_{\rightarrow T}^0$}
\label{Fig:6}
\end{minipage}
\end{figure}

\begin{figure}[h]
\begin{minipage}{0.5\textwidth}
\centerline{
\includegraphics[width=0.76\textwidth]{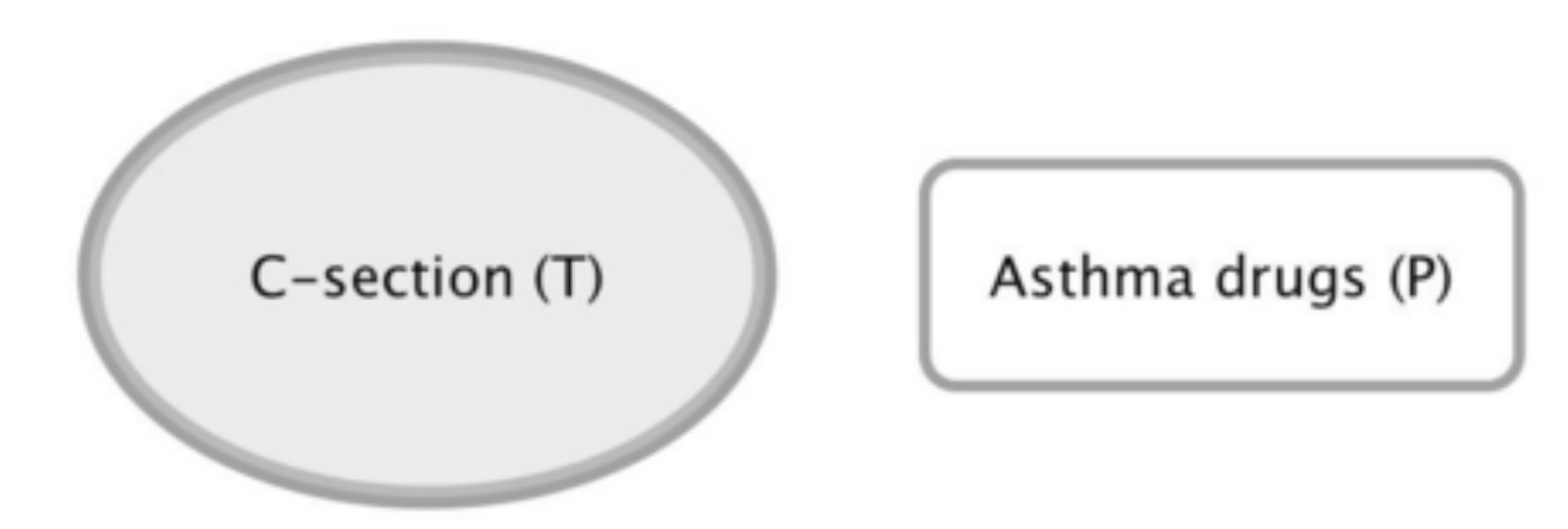}}
\captionsetup{textfont=normal, labelsep=period}
\caption{DAG resulting from MMHC, $ \widehat{Z}_{\rightarrow Y}^1$}
\label{Fig:9}
\end{minipage}
\begin{minipage}{0.5\textwidth}
\centerline{
\includegraphics[width=0.76\textwidth]{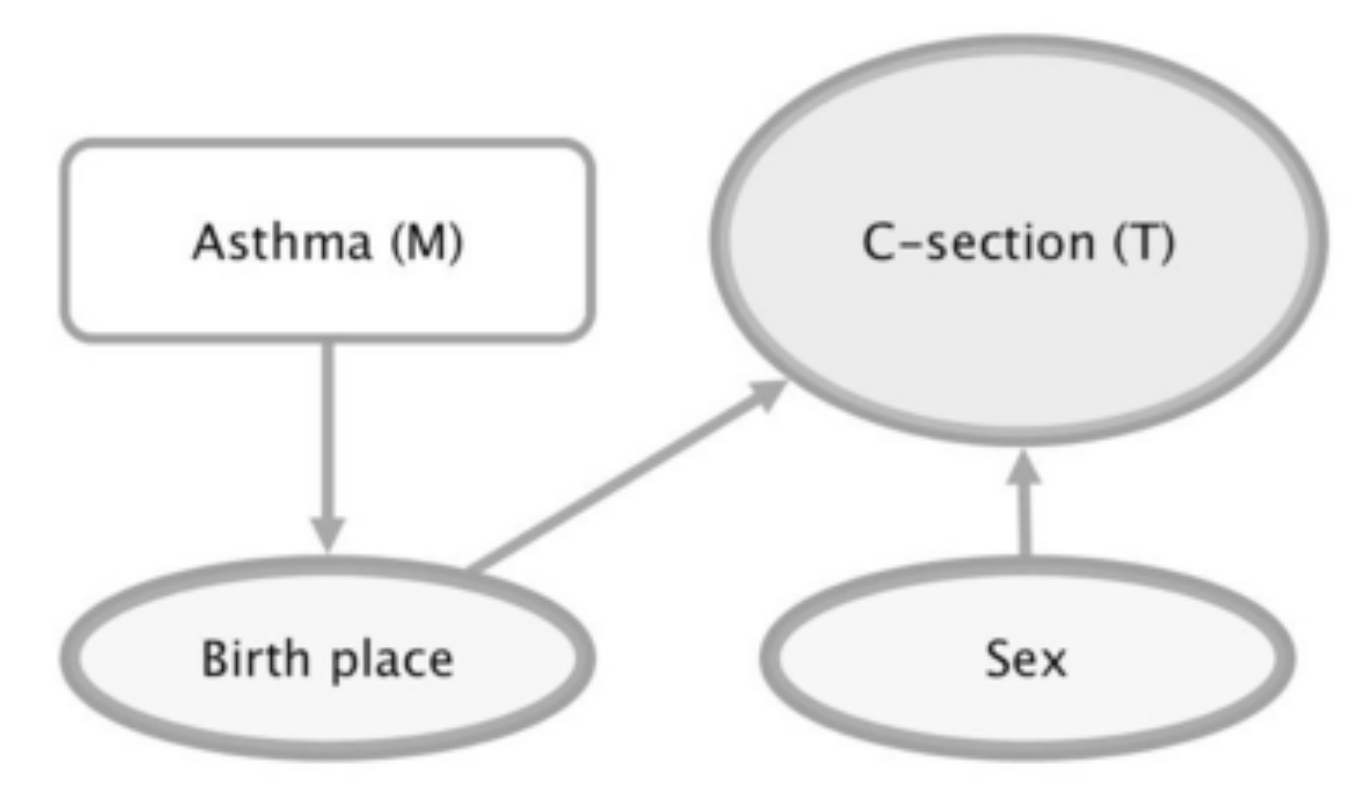}}
\captionsetup{textfont=normal, labelsep=period}
\caption{DAG resulting from MMHC, $ \widehat{Z}_{\rightarrow Y}^0$}
\label{Fig:10}
\end{minipage}
\end{figure}

\end{document}